\documentclass[aos,preprint]{imsart}

\RequirePackage{amsthm,amsmath,amsfonts,amssymb}
\RequirePackage[authoryear]{natbib}
\RequirePackage[colorlinks,citecolor=blue,urlcolor=blue]{hyperref}
\RequirePackage{graphicx}

\startlocaldefs
\theoremstyle{plain}

\newtheorem{theorem}{Theorem}[section]
\newtheorem{lemma}{Lemma}[section]
\newtheorem{proposition}{Proposition}[section]

\newtheorem{remark}{Remark}[section]
\newtheorem{assumption}{Assumption}
\newtheorem{example}{Example}[section]

\theoremstyle{remark}
\newtheorem{definition}[theorem]{Definition}

\endlocaldefs

\usepackage{amsmath, amsfonts, amssymb, amsthm, amstext, graphicx, epsfig, mathtools, enumerate,rotating,verbatim,subfigure}
\usepackage{algorithm,algorithmic}
\usepackage{float}
\usepackage{prettyref,soul}

\usepackage{tabulary}

\usepackage[usenames]{color}
\usepackage{multirow}
\usepackage{tabularx}
\usepackage{booktabs}
\usepackage{hyperref}
\usepackage{bm}
\usepackage{url}
\usepackage{bbding}
\usepackage{mathrsfs}

\usepackage[normalem]{ulem}  
\usepackage[table]{xcolor}  
\usepackage{xcolor}  

\newrefformat{eq}{(\ref{#1})}
\newrefformat{chap}{Chapter~\ref{#1}}
\newrefformat{sec}{Section~\ref{#1}}
\newrefformat{algo}{Algorithm~\ref{#1}}
\newrefformat{fig}{Fig.~\ref{#1}}
\newrefformat{tab}{Table~\ref{#1}}
\newrefformat{rmk}{Remark~\ref{#1}}
\newrefformat{clm}{Claim~\ref{#1}}
\newrefformat{def}{Definition~\ref{#1}}
\newrefformat{cor}{Corollary~\ref{#1}}
\newrefformat{lmm}{Lemma~\ref{#1}}
\newrefformat{lemma}{Lemma~\ref{#1}}
\newrefformat{prop}{Proposition~\ref{#1}}
\newrefformat{app}{Appendix~\ref{#1}}
\newrefformat{ex}{Example~\ref{#1}}
\newrefformat{exer}{Exercise~\ref{#1}}
\newrefformat{soln}{Solution~\ref{#1}}
\newrefformat{cond}{Condition~\ref{#1}}

\def\ep{\textsf{E}}

\def\text#1{\mbox{\rm #1}}

\newcommand{\cov}{{\rm Cov}}

\newcommand{\var}{\mathrm{Var}}

\newcommand{\wh}{\hat}
\renewcommand{\hat}{\widehat}

\renewcommand{\tilde}{\widetilde}

\def\be{\begin{equation}}
\def\ee{\end{equation}}

\def\ep{\mathbb{E}}

\newcommand{\Prob}{\mathbb{P}}
\newcommand{\Expect}{\mathbb{E}}

\definecolor{fire}{rgb}{.86, .08, .24}
\definecolor{airforceblue}{rgb}{0.36, 0.54, 0.86}
\definecolor{amber(sae/ece)}{rgb}{1.0, 0.49, 0.0}
\definecolor{asparagus}{rgb}{0.53, 0.66, 0.42}
\definecolor{darklavender}{rgb}{0.45, 0.31, 0.59}
\definecolor{darksalmon}{rgb}{0.91, 0.59, 0.18}
\definecolor{jasper}{rgb}{0.84, 0.23, 0.24}
\definecolor{blush}{rgb}{0.97, 0.56, 0.81}
\definecolor{cadetgrey}{rgb}{0.57, 0.64, 0.69}

\graphicspath{{main_figs/}}

\usepackage{tikz}
\usepackage{booktabs}
\numberwithin{equation}{section}

\begin{document}

\begin{frontmatter}
\title{Optimal nonparametric inference on network effects with dependent edges}
\runtitle{Testing network effects with dependent edges}

\begin{aug}
\author[A]{\fnms{Wenqin}~\snm{Du}\ead[label=e1]{wenqindu@colostate.edu }},
\author[B]{\fnms{Yuan}~\snm{Zhang}\ead[label=e2]{yzhanghf@stat.osu.edu }},
\and
\author[A]{\fnms{Wen}~\snm{Zhou}\ead[label=e3]{riczw@rams.colostate.edu}}
\address[A]{Department of Statistics, 
Colorado State University\printead[presep={,\ }]{e1,e3}}

\address[B]{Department of Statistics, 
The Ohio State University\printead[presep={,\ }]{e2}}
\end{aug}

\begin{abstract} 
Testing network effects in weighted directed networks is a foundational problem in econometrics, sociology, and psychology. Yet, the prevalent edge dependency poses a significant methodological challenge. Most existing methods are model-based and come with stringent assumptions, limiting their applicability. In response, we introduce a novel, fully nonparametric framework that requires only minimal regularity assumptions. While inspired by recent developments in U-statistic literature \citep{Chen2019U, Zhang2022edgeworth}, our approach notably broadens their scopes. Specifically, we identified and carefully addressed the challenge of  indeterminate degeneracy in the test statistics -- a problem that aforementioned tools do not handle. We established Berry-Esseen type bound for the accuracy of type-I error rate control. Using original analysis, we also proved the minimax optimality of our test's power. Simulations underscore the superiority of our method in computation speed, accuracy, and numerical robustness compared to competing methods. 
We also applied our method to the U.S. faculty hiring network data and discovered intriguing findings. 
\end{abstract}



\end{frontmatter}

\section{Introduction}
\label{section::intro}

\subsection{Motivation}

Amidst the rapid advancements of science and technology, network data representing interactions and relationships among agents/units have become pervasive. Analyzing and understanding the edge-dependency structures within networks remains a methodological challenge, attracting considerable attention from a variety of fields. In social psychology, \cite{Warner1979roundRobin} and \cite{Wong1982roundRobin} introduced a \emph{Social Relations Model} (SRM) that features additive Gaussian random effects for such data. Within this framework, \cite{kenny1988interpersonal} explored the relationships among three nodes, which can be interpreted as various edge dependency structures. Analyzing edge dependency in networks is not only scientifically intriguing but also provides crucial insights when assessing if empirical data aligns with presumed network models \citep{fafchamps2007formation,silva2010currency,graham2020dyadic}. Mathematically, these dependency patterns can be characterized as covariance structures between edges \citep{li2002unified,westveld2011mem,koster2014food,Hoff2021Additive}. They can be empirically estimated by examining the frequencies of small subgraphs, or motifs, involving two or three nodes \citep{opsahl2009clustering,westveld2011mem,underwood2020motif}.
    
We start by briefly reviewing several prominent directed network models with dependent edges from the sociology and econometrics literature. Let $E:=\{e_{i,j}\}_{1\leq \{i,j\}\leq n}$ represent the adjacency matrix, where $e_{i,j}$ might not equal $e_{j,i}$, and $e_{i,i}\equiv 0$ for all $i$.
\begin{itemize}
    \item {\bf Example 1:}
    \cite{li2002unified} introduced a \emph{variance component model}:
    \begin{equation} 
        e_{i,j} = \mu + g_i + g_j + s_{i,j} + d_i - d_j + r_{i,j},
        \label{eqn::variance-component-model}
    \end{equation} 
    where $\{(g_i,d_i)\}_{1 \leq i \leq n}\stackrel{\rm i.i.d.}\sim N(\bm{0}, [\sigma^2_g,\sigma_{gd};\sigma_{gd},\sigma^2_d])$, $\{s_{i,j}=s_{j,i}\}_{1\leq i<j\leq n}\stackrel{\rm i.i.d.}\sim N(0, \sigma^2_s)$, and $\{r_{i,j}=-r_{j,i}\}_{1\leq i<j\leq n}\stackrel{\rm i.i.d.}\sim N(0, \sigma^2_r)$.

    \item {\bf Example 2:}
    \citet{Wong1982roundRobin} presented an \emph{additive Gaussian random-effects model}:
    \begin{equation} 
        e_{i,j} = \mu + a_i + b_j + \epsilon_{i,j},
        \label{eqn::social-relations-model}
    \end{equation}
    where $\{(a_1,b_1),\ldots,(a_n,b_n)\} \stackrel{\rm i.i.d.}\sim N(\bm{0}, [\sigma^2_a,\sigma_{ab};\sigma_{ab},\sigma^2_b])$ and $\{(\epsilon_{i,j},\epsilon_{j,i})\}_{1\leq i<j\leq n} \stackrel{\rm i.i.d.}\sim N(\bm{0}, \sigma^2_\epsilon[1,\rho;\rho,1])$.
    In social sciences, $a_i$, $b_j$ and $\epsilon_{i,j}$ are referred to as \emph{actor}, \emph{partner} and \emph{relationship} terms, respectively \citep{Warner1979roundRobin,kenny1988interpersonal}.

    \item {\bf Example 3:}
    \citet{silva2006log,behar2014trade,graham2020dyadic} proposed a \emph{multiplicative model} for economic networks:
    \begin{equation} 
        e_{i,j} = a_i
        \cdot
        b_j
        \cdot
        \epsilon_{i,j},
        \label{eqn::multiplicative-form}
    \end{equation}
    where $a_i$, $b_i$ and $\epsilon_{i,j}$ are defined similarly to Model \eqref{eqn::social-relations-model}. 
\end{itemize}

In all three models, we observe that $\eta_0 := \cov(e_{i,j}, e_{k,l})=0$, whenever $\{i,j\}\cap\{k,\ell\}=\emptyset$. As pointed out by \citet{zhang2017estimating}, it is impossible to consistently estimate $\eta_1 := \mathrm{Var}(e_{i,j})$\footnote{Notice that $\eta_1$ is independent of the indices $(i,j,k)$, thus well-defined. Same for all other $\eta$'s.} under some settings. There are four other $\eta$'s that are both meaningful and estimable. They are called \emph{network effects} \citep{Squartini2013reciprocityNetwork,cranmer2014reciprocityEffect}, denoted by $\eta_2$ through $\eta_5$ and are formally defined in Table \ref{tab::definition-network-effects}.

\begin{table}[h!] 
\label{tab::definition-network-effects}
    \caption {Definitions of network effects ($i,j,k$ are distinct indices).} 
    \centering
    \resizebox{\linewidth}{!}{
    \begin{tabular}{ccccc}\toprule
    Network effect & \textit{Reciprocity} & \textit{Same-sender} & \textit{Same-receiver} & \textit{Sender-receiver} \\ \midrule 
    Definition & $\eta_2 := \cov(e_{i,j},e_{j,i})$ & $\eta_3 := \cov(e_{j,i},e_{j,k})$ & $\eta_4 := \cov(e_{i,j},e_{k,j})$ & $\eta_5 := \cov(e_{i,j},e_{j,k})$ \\ 
    \raisebox{-1.25ex}[0pt]{Illustration} & 
    \begin{tikzpicture}[baseline=-0.5ex]
    \node[draw,circle,fill,inner sep=1.5pt,label=below:$i$] (i) at (0,0) {};
    \node[draw,circle,fill,inner sep=1.5pt,label=below:$j$] (j) at (1,0) {};
    \draw[->,shorten >=2pt,shorten <=2pt,>=stealth, line width=0.7pt] (i) -- (j);
    \draw[->,shorten >=2pt,shorten <=2pt,>=stealth, line width=0.7pt] (j) -- (i);
    \end{tikzpicture} & 
    \begin{tikzpicture}[baseline=-0.5ex]
    \node[draw,circle,fill,inner sep=1.5pt,label=below:$i$] (i) at (0,0) {};
    \node[draw,circle,fill,inner sep=1.5pt,label=below:$j$] (j) at (1,0) {};
    \node[draw,circle,fill,inner sep=1.5pt,label=below:$k$] (k) at (2,0) {};
    \draw[->,shorten >=2pt,shorten <=2pt,>=stealth, line width=0.7pt] (j) -- (i);
    \draw[->,shorten >=2pt,shorten <=2pt,>=stealth, line width=0.7pt] (j) -- (k);
    \end{tikzpicture} & 
    \begin{tikzpicture}[baseline=-0.5ex]
    \node[draw,circle,fill,inner sep=1.5pt,label=below:$i$] (i) at (0,0) {};
    \node[draw,circle,fill,inner sep=1.5pt,label=below:$j$] (j) at (1,0) {};
    \node[draw,circle,fill,inner sep=1.5pt,label=below:$k$] (k) at (2,0) {};
    \draw[->,shorten >=2pt,shorten <=2pt,>=stealth, line width=0.7pt] (k) -- (j);
    \draw[->,shorten >=2pt,shorten <=2pt,>=stealth, line width=0.7pt] (i) -- (j);
    \end{tikzpicture} & 
    \begin{tikzpicture}[baseline=-0.5ex]
    \node[draw,circle,fill,inner sep=1.5pt,label=below:$i$] (i) at (0,0) {};
    \node[draw,circle,fill,inner sep=1.5pt,label=below:$j$] (j) at (1,0) {};
    \node[draw,circle,fill,inner sep=1.5pt,label=below:$k$] (k) at (2,0) {};
    \draw[->,shorten >=2pt,shorten <=2pt,>=stealth, line width=0.7pt] (i) -- (j);
    \draw[->,shorten >=2pt,shorten <=2pt,>=stealth, line width=0.7pt] (j) -- (k);
    \end{tikzpicture} \\ \bottomrule
    \end{tabular}
    }
\end{table}

Despite the distinct formulations of Models \eqref{eqn::variance-component-model}-\eqref{eqn::multiplicative-form}, in each of them, there exists a one-to-one map between $(\eta_2,\eta_3,\eta_4,\eta_5)$ and set of model parameters. Table \ref{tab::cov-structure-variance-component-model} re-expresses $\eta$'s in terms of model parameters, where the bijection is evident. This is a crucial characteristic shared by many other models, such as \citet{moginley1975dissociated} and \citet{Silverman1976weaklyExch}.

\begin{table}[h!]  
\label{tab::cov-structure-variance-component-model} 
    \caption {Relationship between network effects and parameters under Models \eqref{eqn::variance-component-model}-\eqref{eqn::multiplicative-form}.} 
    \setlength\tabcolsep{5pt}
    \centering
    \resizebox{\linewidth}{!}{
    \begin{tabular}{ccccc}\toprule
     & 
    $\eta_2$ & 
    $\eta_3$ & 
    $\eta_4$ & 
    $\eta_5$  \\ \midrule 
    \cite{li2002unified} & $2\sigma^2_g + \sigma^2_s - 2\sigma^2_d - \sigma^2_r$ & $\sigma^2_g + 2\sigma_{gd} + \sigma^2_d$ & $\sigma^2_g - 2\sigma_{gd} + \sigma^2_d$ & $\sigma^2_g - \sigma^2_d$  \\
    \cite{Wong1982roundRobin} & $2\sigma_{ab}+\rho\sigma^2_\epsilon$ & $\sigma^2_a$ & $\sigma^2_b$ & $\sigma_{ab}$ \\
    \cite{silva2006log} &  $(\sigma_{ab}-1)^2(\rho\sigma^2_\epsilon-1)$ & $\sigma^2_a$ & $\sigma^2_b$ & $\sigma_{ab}$ \\  \bottomrule
    \end{tabular}
    }
\end{table}

As a result, the nonparametric inference tools we develop for network effects will address the pressing need for model selection and validation of several key models across genetics \citep{cockerham1977quadratic,motten2000heritability}, social psychology \citep{kenny1984srm,snijders1999srm,koster2014food,gin2020dyadic}, and economics \citep{ward2007persistent,anderson2011gravity,helpman2008trade,chandrasekhar2016cid,graham2020cid}. Quantifying the significance of network effects necessitates formulating statistical tests with theoretical guarantees under minimal model assumptions, where however, the indeterminate degeneracy poses a compelling challenge. Tackling this challenge is the central focus of this paper.

\subsection{Literature review}
\label{subsec::intro::relation-to-the-literature}

Existing literature predominantly features \emph{model-based} approaches. Notable examples include gravity models \citep{tinbergen1962shaping,anderson2011gravity}, conditionally independent dyad models \citep{chandrasekhar2016cid}, mixed-effects models \citep{gelman2006ref,westveld2011mem}, social relations model \citep{Warner1979roundRobin,Wong1982roundRobin}, exponential-family random graph models (ERGM) \citep{frank1986markov,Snijders2002MCMCeponential,Snijders2006eponentialGraphs, hunter2012computational,Schweinberger2020Concentration}, and block models \citep{yuan2021community}. Among these, SRM has been the cornerstone for most network effect analyses. The most prevalent method within this domain leverages analysis of variance (ANOVA) for parameter estimation \citep{Warner1979roundRobin,bond1996round} and model-based inference, particularly in psychological contexts \citep{card2005gender,eisenkraft2010way,kluger2021dyadic,meagher2021intellectual}. However, this approach is not without its pitfalls \citep{snijders1999srm,nestler2016restricted}.  Challenges include deriving standard errors for ANOVA estimators and the stringent assumption of normality for SRM variances and covariances \citep{lashley1997significance}. In practice, discrete or continuous moderator variables are not easily incorporated into the ANOVA estimation approach \citep{kenny2006analysis,ludtke2013general}. While \citet{li2002unified,nestler2016restricted} proposed likelihood-based methods, they often depend on the premise of additive effect models with normal assumptions. Likewise, Bayesian methods for SRM \citep{gill2001statistical,Hoff2005mixedEffects,hoff2011separable,westveld2011mem,ludtke2013general,koster2014food,Hoff2021Additive} require comparably strong modeling assumptions.

While model-based approaches have been predominant in literature, there is a growing interest in non-parametric methods that can offer more flexibility and require fewer assumptions.
There exist scattered works on permutation and subsampling-based tests for network data, particularly for tests regarding multiple networks \citep{simpson2013permutation,van2022comparing} and inference on membership profiles \citep{fan2022simple,fan2023simpleRC}, as well as in causal inference literature \citep{fredrickson2019permutation}. However, under edge dependence, these existing tools cannot separately test each single network effect; instead, they can only test the group null hypothesis that all network effects are zero \citep{onnela2007structure,snijders2011statistical}.
    
As aforementioned, the network effects outlined in Table \ref{tab::definition-network-effects} can also be empirically estimated using network motifs. This is closely related to a series of works on the \emph{network method-of-moments} \citep{borgs2010moments,bickel2011method,Zhang2022edgeworth}. However, the tools currently available are not adequate for handling directed networks, especially those with potential degeneracy \citep{fisher1983correlation,fan1999central,korolyuk2013theory}.

\subsection{Our contributions}
\label{subsec::intro::our-contributions}

Our contributions span four main aspects. First and foremost, we present a unified nonparametric, model-free inference method for network effects. This sets our work apart from the dominant model-based approaches in existing literature. Our method requires only minimal regularity assumptions, as elaborated in Section \ref{section::our-method}. By avoiding potential model mis-specification, we devise a powerful tool for model selection and validation for network data with dependent edges.
    
Second, we provide an in-depth theoretical analysis that not only quantifies the formulas in our method, but also gives us a good understanding of our method's performance. Distinguished from many other papers studying the similar topic \citep{Chen2019U, Zhang2022edgeworth,shao2023U-statistics}, a major theoretical hurdle is the \emph{indeterminate degeneracy} of our test statistics. Despite the vast literature on conventional U-statistics of known degeneracy status \citep{gregory1977large,serfling1980approximation,fan1999central,korolyuk2013theory,han2018inference,drton2020high}, the indeterminate degeneracy case remains notably under-explored. As pointed out by \citet{shao2023U-statistics}, existing methods for handling indeterminate degeneracy, such as \cite{ho2006two} and \cite{fisher1983correlation}, typically fail to utilize most of available data by extravagant subsampling. Furthermore, in the network setting, we need to deal with ``noisy U-statistics'' \citep{Zhang2022edgeworth} rather than the conventional ``noiseless'' U-statistics, due to observational errors on each edge. In this paper, we devise a diagnostic test to detect the first order degeneracy and tailor downstream inference steps for different situations accordingly. Specifically, to handle the degenerate case, we adapt the \emph{U-statistic reduction} technique \citep{weber1981incomplete, Chen2019U, shao2023U-statistics} that not only reinstates asymptotic normality, but also speeds up computation. U-statistic reduction is a much more refined technique compared to the subsampling approach in \cite{ho2006two} and shows significant advantage in preserving test power. Additionally, we establish the first finite-sample approximation error bounds for network data. These progresses set our work apart from all existing works on network method-of-moments, which exclusively focused on the non-degenerate case.
    
Third, the current literature provides very limited insights into lower bound results about the inference on network effects. Existing lower bound results in network analysis, such as \cite{chao2015rate} and \cite{zhang2016minimax}, typically concentrate on very different topics such as graphon estimation or community detection. Meanwhile, a related work, \cite{shao2022higher} presented a lower bound for two-sample hypothesis testing in the context of comparing two networks.
In this paper, we establish the first set of lower bound results for testing network effects. 
Extending the traditional procedure outlined in works such as \cite{cai2013optimal} and \cite{shao2022higher}, the key innovation in our analysis is the new construction of least-favorable configurations that accounts for the diminishing network effects. In combination with our upper bound results, we show that our method is \emph{nearly rate-optimal} under various scenarios.

Last but not least, we develop and disseminate user-friendly software tailored for practitioners. Our algorithm is computationally efficient and memory-parsimonious, allowing it to easily scale up to networks of over $10^5$ nodes. In Section \ref{section::simulation}, we showcase the advantages of our method in terms of speed, memory efficiency, and numerical robustness when compared to other methods on both synthetic and real-world datasets.

\subsection{Reference chart for main results}

For readers' convenience, we assembled a concise summary of our main results in Table \ref{tab::smry_test}. 
In this table, $\lambda\in[1,2)$ is a user-chosen parameter, with its definition provided in Section \ref{subsec::moment-based-test-of-network-effects::test-same-sender-effect}. Key quantities $\hat\eta$, $\hat\xi$, and $\hat\sigma$ will be defined and discussed in detail in Sections \ref{section::our-method} through \ref{subsec::moment-based-test-of-network-effects::test-reciprocity-effect}.  

\begin{table}[h!]  
\label{tab::smry_test} 
    \caption {Summary of main results.}         
    \centering
    \resizebox{\linewidth}{!}{
    \begin{tabular}{cccccc} \toprule
    Null hypothesis             & Degeneracy      & Test Statistic & Decision & Section & Guarantees \\ \midrule
    $\eta_3=0$                  &     Always  degenerate     & $n^{\lambda/2}~ \wh\eta_{3,J}/\wh\sigma_{3,J}$         & Equation \eqref{test_eta3_test_decomp}  &\ref{subsec::moment-based-test-of-network-effects::test-same-sender-effect}& Theorem \ref{thm_eta3_imcp_concen}    \\
    $\eta_4=0$                  &  Always  degenerate    & $n^{\lambda/2}~ \wh\eta_{4,J}/\wh\sigma_{4,J}$         & Equation \eqref{test_eta4_test_decomp}  &\ref{subsec::moment-based-test-of-network-effects::test-same-receiver-effect}& Theorem B.2    \\ \midrule
    \multirow{2}{*}{$\eta_5=0$} & If non-degenerate & $\sqrt{n}\wh\eta_{5,n}/\wh\xi_{5,1}$        & Equation \eqref{test_eta5_test} &\ref{subsubsec::moment-based-test-of-network-effects::test-sender-receiver-effect::non-degenerate-case}& Theorem \ref{thm_var5_concen}    \\
        & If degenerate    & $n^{\lambda/2}~ \wh\eta_{5,J}/\wh\sigma_{5,J}$        & Equation \eqref{test_eta5_test_decomp} &\ref{subsubsec::moment-based-test-of-network-effects::test-sender-receiver-effect::degenerate-case}& Theorem \ref{thm_eta5_imcp_concen}   \\ \midrule
    \multirow{2}{*}{$\eta_2=0$} & If non-degenerate & $\sqrt{n}\wh\eta_{2,n}/\wh\xi_{2,1}$        & Equation \eqref{test_eta2_test} &\ref{subsubsec::moment-based-test-of-network-effects::test-reciprocity-effect::non-degenerate-case}& Theorem \ref{thm_var2_concen}   \\
        & If degenerate    & $n^{\lambda/2}~ \wh\eta_{2,J}/\wh\sigma_{2,J}$        & Equation \eqref{test_eta2_test_decomp} &\ref{subsubsec::moment-based-test-of-network-effects::test-reciprocity-effect::degenerate-case}& Theorem \ref{thm_eta2_imcp_concen}   \\ \bottomrule
    \end{tabular}
    }
\end{table}

\section{Problem set up: network effects in exchangeable networks}
\label{section::network-effects-in-exchangeable-networks}

In this section, we first formally set up the problem. 
Recall that in Section \ref{section::intro}, we described network effects under Models \eqref{eqn::variance-component-model}-\eqref{eqn::multiplicative-form}. It turns out that these models are all special cases of a much larger family of models, namely, \emph{exchangeable networks}.

\begin{definition}[Exchangeable networks, \cite{hoff2007modeling}]
\label{def::WEdef} 
    A network $E:=\{e_{i,j}\}_{1\leq \{i,j\}\leq n}$ is \emph{exchangeable}, if it is embedded in an exchangeable infinite matrix $\{e_{i,j}\}_{i,j\in \mathbb{N}^+}$, where for any node permutation $\pi:  \mathbb{N}^+\leftrightarrow \mathbb{N}^+$, we have
    $\{e_{i,j}\}_{i,j\in \mathbb{N}^+} \stackrel{d}=\{e_{\pi(i),\pi(j)}\}_{i,j\in \mathbb{N}^+}$. 
\end{definition}

Node exchangeability implies that the order in which nodes are collected does not carry any information. Importantly, exchangeability leads to a succinct and elegant universal representation, called the \emph{Aldous-Hoover Representation}, encompassing a broad family of network models. These include SRM \citep{Warner1979roundRobin,Wong1982roundRobin,kenny1984srm,snijders1999srm}, mixed effects models \citep{gelman2006ref,westveld2011mem}, conditionally independent dyad models \citep{chandrasekhar2016cid,graham2020cid}, alongside row-column exchangeable models \citep{aldous1985exchangeability}, and crossed random effect models \citep{Art2012a}. The graphon model delineates an i.i.d. procedure for sampling nodes from a hyper-population of ``all kinds of nodes''.

\begin{theorem}[\cite{hoover1979relations,aldous1981representations}] 
\label{exch_2}
    A network $\{e_{i,j}\}_{1\leq \{i,j\}\leq n}$ satisfying Definition \ref{def::WEdef} admits the universal representation
    \begin{align}\label{eqn::Aldous-Hoover-representation}
    \{e_{i,j}\}_{1\leq \{i,j\}\leq n}
    \stackrel{d}=&~
    \{F(X_i,X_j,X_{(i,j)})\}_{1\leq \{i,j\}\leq n},
    \end{align}
    where $F$ is a latent, potentially asymmetric function that encodes all network structures, and latent variables are generated as $\{X_i\}_{1\leq i\leq n} \cup \{X_{(i,j)} = X_{(j,i)}\}_{1\leq i<j\leq n}\stackrel{\rm i.i.d.}\sim \mathrm{Uniform}[0,1]$.
\end{theorem}

Representation \eqref{eqn::Aldous-Hoover-representation} ensures that the network effects $\eta_2$ through $\eta_5$ in Table \ref{tab::definition-network-effects} do not depend on the specific values of indices $i,j$ and $k$, thus are all well-defined. The primary objective of our paper is to test the hypotheses
\begin{align}
\label{nulls}
    H_0^{(\ell)}:\eta_\ell=0,
    \quad\textrm{versus}\quad
    H_a^{(\ell)}:\eta_\ell\neq0,
    \quad \textrm{for }\ell=2,3,4,5.
\end{align}
For technical reasons, we will present our tests for these $\eta$'s in a different order than the one in which they were initially introduced. We will first study $\eta_3$ and $\eta_4$, then $\eta_5$, and lastly $\eta_2$.

\section{Inference for same-sender effect \texorpdfstring{$\eta_3$}{eta3} and same-receiver effect \texorpdfstring{$\eta_4$}{eta4}}
\label{section::our-method}

The testing procedures for all $\eta$'s exhibit some similarities, yet there are also notable distinctions. We start with designing and analyzing the test for $\eta_3$ in the greatest detail, serving as an anchor point for our presentation.

\subsection{Inference for $\eta_3$}
\label{subsec::moment-based-test-of-network-effects::test-same-sender-effect} 

By definition,
\begin{align}
    \eta_3 
    :=&~ 
    \cov(e_{i,j},e_{i,k}) = \Expect(e_{i,j}e_{i,k}) - \Expect (e_{i,j}) \Expect (e_{i,k})= \Expect(e_{i,j}e_{i,k}) - \mu_e^2,
    \label{eqn::eta_3-temp-1}
\end{align}
where $\mu_e:=\Expect (e_{i,j})=\Expect (e_{i,k})$. 
To reduce symbols, we introduce the shorthand
$S_3 := \{
        \{(i,j), (i,k)\}: \textrm{ distinct }i, j, k \in [n]:=\{1,2,\ldots,n\}
    \}$
to represent all edge pairs originating from the same sender and towards different receivers. 
Naturally, this notation allows us to devise the estimator $\hat\eta_{3,n}$ as the empirical counterpart to \eqref{eqn::eta_3-temp-1}
\begin{align}
    \wh \eta_{3,n} 
    :=&~ 
    |S_3|^{-1} 
    \sum_{\{(i,j), (i,k)\}\in S_3} 
    e_{i,j} e_{i,k} 
    - 
    \Big\{
        n^{-1}(n-1)^{-1} \sum^n_{i\neq j} e_{i,j} 
    \Big\}^2,
    \label{eta_3mmt}
\end{align}
where in the second term on the RHS of \eqref{eta_3mmt}, we used a plug-in estimator for $\mu^2_e$ to reduce computation while introducing only a negligible bias.
    
Our next step is to re-express $\hat \eta_{3,n}$ in terms of network moments, as elaborated in \citet{Zhang2022edgeworth}.
Here, the \emph{sample network moment} indexed by a motif involving 
$r$ nodes is
\begin{equation}
    \wh U_n := {n \choose r}^{-1} \sum_{1 \leq i_1<\ldots<i_r \leq n} h(E_{i_1,\ldots,i_r}),
    \label{netU}
\end{equation}
where $E_{i_1,\ldots,i_r}$ is the induced sub-network of $E$ between nodes $\{i_1,\ldots,i_r\}$. The network moment in \eqref{netU} is also referred to as \emph{network U-statistic} \citep{shao2023U-statistics}, especially when compared to conventional U-statistics. The key idea is to decompose $e_{i,j}$ into two parts \citep{Zhang2022edgeworth}, as follows:
\begin{equation}
    e_{i,j} = f(X_i,X_j) + \rho_{i,j},
    \label{e_decmp}
\end{equation}
where $f(X_i,X_j) := \Expect(e_{i,j}|X_i,X_j) = \Expect\{F(X_i, X_j, X_{(i,j)})|X_i,X_j\}$ encodes the random variation solely due to $(X_i,X_j)$, and $\rho_{i,j} := e_{i,j}- f(X_i,X_j)$ captures the observational error conditional on $(X_i,X_j)$. Applying \eqref{e_decmp}, we can rewrite \eqref{netU} as
\begin{align}
    \wh U_n 
    :=&~ 
    {n \choose r}^{-1} \sum_{1 \leq i_1<\ldots<i_r \leq n} h(\tilde E_{i_1,\ldots,i_r}) + R_n,
    \label{decp_netU}
\end{align}
where $\tilde E_{i_1,\ldots,i_r}$ is the induced sub-matrix of of the expectation matrix 
\begin{align}
    \tilde E := \ep[E|X_1,\ldots,X_n] = \{f(X_i,X_j)\}_{1\leq \{i,j\}\leq n},
    \label{equation::tE-def}
\end{align}
and the order of remainder $R_n$ will be specified in Proposition \ref{degen_remain_eta3} later. 
The decomposition in \eqref{decp_netU} is particularly useful for analysis, as its first term is a conventional ``noiseless'' U-statistic with inputs $\{X_i\}_{1\leq i\leq n}$. This characteristic simplifies the subsequent analysis.

\subsubsection{Degeneracy status of $\wh\eta_{3,n}$}
\label{subsubsection::Nontrivial-degeneracy-of-eta3}

Now, we return to the analysis of \eqref{eta_3mmt}. We make the following mild assumption about the exchangeable network in Definition \ref{def::WEdef}.

\begin{assumption}
\label{assumption::sub-exp}
    For any $(i,j)$ pair, assume $e_{i,j}, f(X_i,X_j)$, $\Expect(e_{i,j}|X_i)$, and $\Expect(e_{j,i}|X_i)$ are all sub-exponential random variables \citep{vershynin2018high}. 
    Additionally, for all edge pair $(e_{i_1,j_1},e_{i_2,j_2})$, where $i_1,i_2,j_1,j_2 \in \{i,j,k: (i,j,k)\textrm{ distinct}\}$, assume $\Expect(e_{i_1,j_1}e_{i_2,j_2}|X_i)$ is sub-exponential.
\end{assumption}
    
With the shorthand $h_1(E_{i,j}) =  (e_{i,j} + e_{j,i})/2$ and $h_3(E_{i,j,k}) = (e_{i,j} e_{i,k} + e_{j,i} e_{j,k} + e_{k,i} e_{k,j})/3$, we can further express \eqref{eta_3mmt} using two \emph{network moments} as follows:
\begin{equation}
\label{eta3_networkmmts}
    \wh\eta_{3,n} = {n \choose 3}^{-1} \sum_{1 \leq i<j<k \leq n} h_3(E_{i,j,k}) - \Big\{{n \choose 2}^{-1} \sum_{1 \leq i<j \leq n} h_1(E_{i,j}) \Big\}^2.
\end{equation}
It turns out that the stochastic variation in $\hat \eta_{3,n}$ is dominated by the randomness due to $X_i$'s. With \eqref{equation::tE-def} we can replace $E_{i,j,k}$ by $\tilde E_{i,j,k}$ and $E_{i,j}$ by $\tilde E_{i,j}$ in \eqref{eta3_networkmmts} and formulate the main part of $\hat \eta_{3,n}$ as
\begin{equation}
\label{H3_def}
    H_{3,n} = {n \choose 3}^{-1} \sum_{1 \leq i<j<k \leq n} h_3(\tilde E_{i,j,k}) - \Big\{{n \choose 2}^{-1} \sum_{1 \leq i<j \leq n} h_1(\tilde E_{i,j}) \Big\}^2,
\end{equation}
In the decomposition $\wh\eta_{3,n}=H_{3,n} + (\wh\eta_{3,n} - H_{3,n})$, we have the following result.
\begin{proposition} 
\label{degen_remain_eta3}
    Consider an exchangeable network which admits the universal representation in Theorem \ref{exch_2}. Under Assumption \ref{assumption::sub-exp}, we have $\wh\eta_{3,n} - H_{3,n}=\tilde O_p(n^{-1}\log n)$, where we write $Y_n = \tilde O_p(\alpha_n)$ if $\Prob(|Y_n| \geq C\alpha_n) < n^{-1}$ for some constant $C>0$.
\end{proposition}
This guides us to design the variance estimation for $\wh \eta_{3,n}$ by studying $H_{3,n}$. Denote the two terms in the expression of $H_{3,n}$ as $U_{3,n} = {n \choose 3}^{-1} \sum_{1 \leq i<j<k \leq n} h_3(\tilde E_{i,j,k})$ and $U_{1,n} = {n \choose 2}^{-1} \sum_{1 \leq i<j \leq n} h_1(\tilde E_{i,j})$. As a conventional U-statistic, $U_{3,n}$ admits the following Hoeffding's decomposition. Recall $\mu_e$ in \eqref{eqn::eta_3-temp-1}, 
\begin{align}
    U_{3,n} - \eta_3 - \mu^2_e
    =&~ 
    \frac{3}{n}\sum_{1 \leq i \leq n} g_{3,1}(X_i)
    + 
    \frac{6}{n(n-1)}\sum_{1 \leq i<j \leq n}g_{3,2}(X_i,X_j) 
    + 
    \tilde O_p(n^{-3/2}\log^{3/2} n),
    \label{U_3n}
\end{align}
where $g_{3,1}(x_i) := \Expect\{h_3(\tilde E_{i,j,k})|X_i=x_i\} - \Expect(e_{i,j}e_{i,k})$, and $g_{3,2}(x_i,x_j) := \Expect\{h_3(\tilde E_{i,j,k})|X_i=x_i,X_j=x_j\} - g_{3,1}(x_i) - g_{3,1}(x_j) - \Expect(e_{i,j}e_{i,k})$. Similarly, $U_{1,n}$ admits
\begin{align}
    U_{1,n} - \mu_e
    =&~ 
    \frac{2}{n}\sum_{1 \leq i \leq n} g_{1,1}(X_i)
    + 
    \frac{2}{n(n-1)}\sum_{1 \leq i<j \leq n}g_{1,2}(X_i,X_j)
    +
    \tilde O_p(n^{-3/2}\log^{3/2} n),
    \label{U_1n}
\end{align}
where we set $g_{1,1}(x_i) := \Expect\{h_1(\tilde E_{i,j})|X_i=x_i\} - \mu_e$ and $g_{1,2}(x_i,x_j) := \Expect\{h_1(\tilde E_{i,j})|X_i=x_i,X_j=x_j\} - g_{1,1}(x_i) - g_{1,1}(x_j) - \mu_e$. We emphasize that $g_{1,1}(\cdot)$, $g_{1,2}(\cdot),g_{3,1}(\cdot)$, and $g_{3,2}(\cdot)$ serve merely as intermediate shorthand to streamline the narrative. Our analysis will solely concentrate on $g_{\eta_3,1}(X_i)$ and $g_{\eta_3,2}(X_i,X_j)$. Combining \eqref{H3_def}-\eqref{U_1n} and reorganizing terms lead to 
\begin{align}
    H_{3,n} - \eta_3  
    =&~ \frac{1}{n}\sum_{1 \leq i \leq n} g_{\eta_3,1}(X_i)+ {n \choose 2}^{-1}\sum_{1 \leq i<j \leq n}g_{\eta_3,2}(X_i,X_j) 
    + 4n^{-1}\Expect\{g^2_{1,1}(X_1)\} + R_{H_3,n},
    \label{H_3n}
\end{align}
where $g_{\eta_3,1}(X_i) := 3g_{3,1}(X_i) - 4\mu_e g_{1,1}(X_i)$, from which we define 
\begin{align}
\label{equation::xi31-def}
    \xi_{3,1} := \{\var(g_{\eta_3,1}(X_1))\}^{-1/2};       
\end{align}
and $g_{\eta_3,2}(X_i,X_j) := 3g_{3,2}(X_i,X_j) - 2\mu_e g_{1,2}(X_i,X_j) - 4n^{-1}(n-1)g_{1,1}(X_i)g_{1,1}(X_j)$, and all $g_{\eta_3,k}(X_{i_1},\ldots, X_{i_k})$ terms are mutually uncorrelated. The remainder term $R_{H_3,n}$ collecting all other terms is characterized in Lemma \ref{lemma_eta3_remainder}.
\begin{lemma}
\label{lemma_eta3_remainder}
    Consider an exchangeable network which admits the universal representation in Theorem \ref{exch_2}. Under Assumption \ref{assumption::sub-exp}, the remainder in \eqref{H_3n} satisfies $R_{H_3,n} = \tilde O_p(n^{-3/2}\log^{3/2}n)$.
\end{lemma}

We can define the \emph{degeneracy} of the estimator $\hat\eta_{3,n}$ using \eqref{equation::xi31-def}.
It is considered \textit{degenerate} if $\xi_{3,1} \equiv 0$ and \emph{non-degenerate} if $\xi_{3,1} \geq {\rm Constant}>0$.
Next, we report an interesting and important discovery: the value of $\eta_3$ crucially impacts the degeneracy status of $\hat \eta_{3,n}$. It seems that when $\xi_{3,1} \geq {\rm Constant}>0$, \eqref{H_3n} would reduce to the non-degenerate noisy U-statistic setting in \citet{Zhang2022edgeworth}. However, we find that $H_0^{(3)}$ in fact rules out this case.

\begin{proposition} 
\label{prop_eta3}
    The null hypothesis $H_0^{(3)}: \eta_3=0$ implies that $\xi_{3,1} \equiv 0$. On the other hand, if $\xi_{3,1} \geq \mathrm{Constant}> 0$, then $|\eta_3| \geq \mathrm{Constant}> 0$. 
\end{proposition}

Under $H_0^{(3)}$, we have $3g_{3,1}(X_i) = 4\mu_e g_{1,1}(X_i) = 2\mu_e\Expect\{f(X_j,X_i)|X_i\} - 2\mu_e^2$, which, along with the fact $g_{\eta_3,1}(X_i) = 3g_{3,1}(X_i) - 4\mu_e g_{1,1}(X_i)$, yields Proposition \ref{prop_eta3}. We refer to Section  A.1 in the Supplementary Materials for the detailed proof of Proposition \ref{prop_eta3}. Proposition \ref{prop_eta3} implies the commonly observed first order degeneracy of U-statistics under the null hypothesis \citep{serfling1980approximation,drton2020high}. However, the degree of degeneracy of $\hat\eta_{3,n}$ can exceed order $1$. In fact, under some network generation mechanisms, we have $\var(g_{\eta_3,2}(X_1,X_2)) \geq \textrm{Constant}>0$, while for others, $\var(g_{\eta_3,2}(X_1,X_2))$ may diminish to $0$ as $n\to\infty$ or exactly equals to $0$. This highlights the challenge posed by the indeterminate degeneracy of network U-statistics, as discussed in Section \ref{subsec::intro::our-contributions}, which is exemplified by the following concrete cases.

\begin{example}[Indeterminate degeneracy of $\hat\eta_{3,n}$ under $H_0^{(3)}$] 
\label{example::eta3-indeterminant-degen-g2}
    Inherit notation from Theorem \ref{exch_2}. Additionally, set $\{\epsilon_{i,j}\}_{1 \leq \{i,j\} \leq n}$ to be i.i.d., mean-zero random variables, independent of everything else. Then, 
    \begin{enumerate}[(i)]
        \item if we model $e_{i,j}=X_{(i,j)}+\epsilon_{i,j}$, then $g_{\eta_3,2}(X_i,X_j) \equiv 0$; \label{eta3-1}
        \item if we model $e_{i,j}=X_j+\epsilon_{i,j}$, then $g_{\eta_3,2}(X_i,X_j)= n^{-1}(X_i-1/2)(X_j-1/2)$, thus $\var(g_{\eta_3,2}(X_i,X_j))=O(n^{-2})$; \label{eta3-2}
        \item if we model $e_{i,j}=(X_i-1/2)(X_j-1/2)+\epsilon_{i,j}$, then $g_{\eta_3,2}(X_i,X_j)= (X_i-1/2) (X_j-1/2)/12$, thus $\var(g_{\eta_3,2}(X_i,X_j))\geq\mathrm{Constant}>0$. \label{eta3-3}
    \end{enumerate}
\end{example}
    
The asymptotic distribution of $\wh \eta_{3,n}$ under $H_0^{(3)}$ depends on its degeneracy status and $\wh\eta_{3,n} - H_{3,n}$. When $\var(g_{\eta_3,2})\geq {\rm Constant}>0$, the limiting distribution of $H_{3,n}$ is a mixture of $\chi^2$ distributions. Otherwise, it becomes a complicated \emph{Gaussian chaos distribution} \citep{van2000asymptotic}. As discussed in Section \ref{subsec::intro::our-contributions}, existing methods do not address the indeterminate degeneracy of U-statistics. Moreover, in view of Proposition \ref{degen_remain_eta3}, $\wh\eta_{3,n} - H_{3,n}$ may either be on the same order with the quadratic part in the decomposition of $H_{3,n}$, which is the second component in \eqref{H_3n} (see \eqref{eta3-3} in Example \ref{example::eta3-indeterminant-degen-g2}); or dominate it (see \eqref{eta3-1} and \eqref{eta3-2} in Example \ref{example::eta3-indeterminant-degen-g2}). This further exacerbates the difficulty of deriving the limiting distribution of $\wh \eta_{3,n}$.

\subsubsection{Testing $\eta_{3}$ using the U-statistics reduction for $\wh\eta_{3,n}$}
\label{subsubsection::Testing-using-the-U-statistics-reduction} 

Fortunately, we can employ the trick of \emph{U-statistic reduction} to reinstate asymptotic normality and accelerate computation \citep{weber1981incomplete, Chen2019U, shao2023U-statistics}. Some algebra shows that $\wh\eta_{3,n}$ could be rewritten as 
\begin{align}
    \wh\eta_{3,n} = {n \choose 4}^{-1} \sum_{1 \leq i<j<k<l \leq n} \psi_3(E_{i,j,k,l}),
    \label{equation::eta_3n_reform}
\end{align}
where $\psi_3(E_{i,j,k,l}) := \sum\nolimits_{\{i_1,i_2,i_3\}\in\{i,j,k,l\}}h_3(E_{i_1,i_2,i_3})/4 - h_6(E_{i,j,k,l}) + r(E_{i,j,k,l})$ is a random variable of constant order in view of Assumption \ref{assumption::sub-exp}. Here, we define the shorthand
\begin{align*}
    h_6(E_{i,j,k,l}):=\frac{1}{4!}\sum\nolimits^{\{i,j,k,l\}}_{\{i_1,i_2,i_3,i_4\}} e_{i_1,i_2}e_{i_3,i_4},
\end{align*}
where $\sum\nolimits^{\{i,j,k,l\}}_{\{i_1, \ldots,i_m\}}$ is summing over all permutations of the $m$-tuples ($m\leq 4$) of indices $\{i,j,k,l\}$, as defined in \citet{yao2018testing}. Also, $r(E_{i,j,k,l}) = \sum\nolimits^{\{i,j,k,l\}}_{\{i_1,i_2\}} (e_{i_1,i_2}e_{i_2,i_1} +   e_{i_1,i_2}^2) / \{12 (n^2-n)\} + (n-2) \sum\nolimits^{\{i,j,k,l\}}_{\{i_1,i_2,i_3\}}  \{h_3(E_{i_1,i_2,i_3}) + h_4(E_{i_1,i_2,i_3}) + 2 h_5(E_{i_1,i_2,i_3})\} / \{4 (n^2 - n)\} + (6-4n) h_6(E_{i,j,k,l}) / (n^2-n)$, where $h_4(E_{i,j,k}) = (e_{i,j} e_{k,j} + e_{j,i} e_{k,i} + e_{i,k} e_{j,k})/3$ and $h_5(E_{i,j,k}) = \sum\nolimits^{\{i,j,k\}}_{\{i_1,i_2,i_3\}} e_{i_1,i_2}e_{i_2,i_3} /6$, within which $\sum\nolimits^{\{i,j,k\}}_{\{i_1, i_2,i_3\}}$ denoting the summing over all permutations of the $3$-tuples of indices $\{i,j,k\}$. It is not hard to show that $r(E_{i,j,k,l})$ is at the order of $n^{-1}$. As a result, $r(E_{i,j,k,l})$ is dominated by the first two terms of $\psi_3(E_{i,j,k,l})$.

In view of the representation in \eqref{equation::eta_3n_reform}, we can consider a reduced version of $\hat \eta_{3,n}$, commonly known as the \emph{reduced network moment}, 
\begin{equation}
    \wh\eta_{3,J} := n^{-\lambda} \sum_{(i,j,k,l) \in J_{n,\lambda}} \psi_3(E_{i,j,k,l}),
\label{incmp_3}
\end{equation}
where $J_{n,\lambda} := (I^{(1)}_4, I^{(2)}_4, ..., I^{(|J_{n,\lambda}|)}_4)$ is a subsample (with replacement) of size $|J_{n,\lambda}| = n^\lambda$ from ${\cal C}_n^4:=\{\textrm{All possible 4-tuples among }[n]\}$ \citep{shao2023U-statistics}, and $\lambda$ controls the level of computation reduction. In the following decomposition 
\begin{align}
    \wh\eta_{3,J} - \eta_3 
    =&~ 
    (\wh\eta_{3,n} - \eta_3) 
    + 
    n^{-\lambda} \sum_{(i,j,k,l) \in J_{n,\lambda}} \{\psi_3(E_{i,j,k,l})-\wh\eta_{3,n}\}:=
    (\wh\eta_{3,n} - \eta_3)
    +
    V_{3,J},
\label{incmp_3_decmp}
\end{align}
the second term on the right-hand side dominates, necessitating the tailoring of our variance estimator accordingly. By Propositions \ref{degen_remain_eta3} and \ref{prop_eta3} and Lemma \ref{lemma_eta3_remainder}, we have $\wh\eta_{3,n} - \eta_3=\tilde O_p(n^{-1}\log n)$. Also, conditioning on $\{e_{i,j}\}_{1\leq \{i,j\} \leq n}$, we can view $V_{3,J}$ as an average of independent mean-zero random variables. Therefore, we obtain $\var(V_{3,J}|\{e_{i,j}\})=n^{-\lambda}\sigma^2_{3,J}=O(n^{-\lambda})$. Hence, we can estimate $\sigma^2_{3,J}$ by 
\begin{align}
    \wh\sigma^2_{3,J}=n^{-\lambda} \sum_{(i,j,k,l) \in J_{n,\lambda}}\{ \psi_3(E_{i,j,k,l}) - \wh\eta_{3,J}\}^2.
\end{align}
This leads to the following studentization
\begin{align}
    (\wh\eta_{3,J} - \eta_3)/(n^{-\lambda/2}\wh\sigma_{3,J}) 
    =&~
    V_{3,J}/(n^{-\lambda/2}\wh\sigma_{3,J})
    +
    \tilde O_p(n^{\lambda/2-1}\log n).
    \label{incmp_3_decmp_studz_estimated_sd}  
\end{align}
By examining the first term on the right-hand side of \eqref{incmp_3_decmp_studz_estimated_sd}, we derive a Berry-Esseen type bound for the studentized estimator, as articulated in Theorem \ref{thm_eta3_imcp_concen_split-1}.

\begin{theorem}
\label{thm_eta3_imcp_concen_split-1} 
    Consider an exchangeable network satisfying Definition \ref{def::WEdef} which admits the universal representation in Theorem \ref{exch_2}. Under Assumption \ref{assumption::sub-exp}, when $\xi_{3,1} \equiv 0$, for $\lambda \in [1,2)$ such that $n^\lambda \in \mathbb{Z}$, we have 
    \begin{align}
    \label{eqn::theorem::eta3}
    \sup_x\Big|\Prob\big[n^{\lambda/2}(\wh\eta_{3,J}-\eta_3)/\wh\sigma_{3,J}\leq x \big]-\Phi(x)\Big|             \leq&~ 
    C(n^{\lambda/2-1}+n^{-\lambda/2})\log n,
    \end{align}
    for some constant $C>0$, where $\Phi(\cdot)$ is the cumulative distribution function of the standard normal distribution. 
\end{theorem}

The first term in the error bound in \eqref{eqn::theorem::eta3} bounds $|\hat\eta_{3,n} - \eta_3|$ in \eqref{incmp_3_decmp}, and the second term accounts for the approximation error in $\wh\sigma_{3,J} - \sigma_{3,J}$. Theorem \ref{thm_eta3_imcp_concen_split-1} extends the results in \citet{Chen2019U} and \cite{shao2023U-statistics}: these prior works focus on conventional U-statistics and non-degenerate network moment statistics, respectively, whereas our Theorem \ref{thm_eta3_imcp_concen_split-1} studies network moment statistics under indeterminate degeneracy. Based on Theorem \ref{thm_eta3_imcp_concen_split-1}, for $\lambda \in [1,2)$ such that $n^\lambda \in \mathbb{Z}$ and any given significance level $\alpha \in (0, 1)$, we can test the hypotheses in \eqref{nulls} as follows.
\begin{equation} 
\label{test_eta3_test_decomp}
    \text{Reject~} H_0^{(3)} \text{~if~~} \big|n^{\lambda/2}~ \wh\eta_{3,J}/\wh\sigma_{3,J}\big| > \Phi^{-1}(1-\alpha/2),
\end{equation}
where $\Phi^{-1}(1-\alpha/2)$ denotes the $1-\alpha/2$ lower quantile of $N(0,1)$.

\subsubsection{Theoretical guarantees}
\label{Sec313}

Using Theorem \ref{thm_eta3_imcp_concen_split-1}, we establish the quantification of finite-sample error controls of test \eqref{test_eta3_test_decomp} in Theorem \ref{thm_eta3_imcp_concen}.  
\begin{theorem}
\label{thm_eta3_imcp_concen} 
    Consider an exchangeable network satisfying Definition \ref{def::WEdef} which admits the universal representation in Theorem \ref{exch_2}. 
    Under Assumption \ref{assumption::sub-exp}, the type-I error rate of test \eqref{test_eta3_test_decomp} is $\alpha + O((n^{\lambda/2-1}+n^{-\lambda/2})\log n)$.
    The type-II error rate satisfies:
    \begin{enumerate}[(i)]
        \item when $\xi_{3,1} \equiv 0$, the type-II error rate of test \eqref{test_eta3_test_decomp} is $o(1)$ if $\eta_3=\omega(n^{-\lambda/2})$;
        \item when $\xi_{3,1} \geq \mathrm{Constant}>0$ (which implies $\eta_3\geq\mathrm{Constant}>0$ according to Proposition \ref{prop_eta3}), the type-II error rate of test \eqref{test_eta3_test_decomp} is $o(1)$.
    \end{enumerate}
\end{theorem}

Theorem \ref{thm_eta3_imcp_concen} quantifies the finite-sample trade-off between the type-I error rate control accuracy and the power of the proposed test. With respect to $\lambda$, the error bound for accurately controlling the type-I error rate is optimized at $\lambda=1$, which yields $\Prob_{H_0^{(3)}}\{\text{Reject~} H_0^{(3)}\} = \alpha + O(n^{-1/2}\log n)$. Although setting a larger $\lambda$ increases the computational complexity of $\wh\eta_{3,J}$, it also enhances the power of test \eqref{test_eta3_test_decomp}. To see this, notice that the minimum separation condition for power consistency $\eta_3=\omega(n^{-\lambda/2})$ is easier to satisfy with a larger $\lambda$ (also see Section \ref{subsection::optimality-for-testing-ets3and4}). Moreover, when $\xi_{3,1} \geq\mathrm{Constant}>0$, the power of our test also grows with $\lambda$. This can be attributed to that the test statistic departs more significantly from zero as $\lambda$ increases, as established in the proof of Theorem \ref{thm_eta3_imcp_concen}.

\subsection{Inference for $\eta_4$}
\label{subsec::moment-based-test-of-network-effects::test-same-receiver-effect}

The definitions of $\eta_3$ and $\eta_4$ are mirror to each other. No surprise that their inference procedures are similar. Therefore, in this part, we provide a highly-compressed description of the testing procedure for $\eta_4$. 
    
Write $h_4(E_{i,j,k}) = (e_{i,j} e_{k,j} + e_{j,i} e_{k,i} + e_{i,k} e_{j,k})/3$. We can estimate $\eta_4$ by $\wh\eta_{4,n} := {n \choose 3}^{-1} \sum_{1 \leq i<j<k \leq n} h_4(E_{i,j,k}) - \{{n \choose 2}^{-1} \sum_{1 \leq i<j \leq n} h_1(E_{i,j}) \}^2$. As exemplified in Example \ref{example::eta3-indeterminant-degen-g2}, we also encounter the challenge of indeterminate degeneracy of $\hat\eta_{4,n}$ under $H_0^{(4)}$. Consequently, we again opt to test $\eta_4$ by utilizing U-statistics reduction, considering the reduced test statistic $\wh\eta_{4,J} := n^{-\lambda} \sum_{(i,j,k,l) \in J_{n,\lambda}} \psi_4(E_{i,j,k,l})$. Here, $\psi_4(E_{i,j,k,l})$ is defined similarly to $\psi_3(E_{i,j,k,l})$ in Section \ref{subsubsection::Testing-using-the-U-statistics-reduction} by replacing $h_3(E_{i_1,i_2,i_3})$ with $h_4(E_{i_1,i_2,i_3})$. We consider the variance estimator $\wh\sigma^2_{4,J}=n^{-\lambda} \sum_{(i,j,k,l) \in J_{n,\lambda}}\{ \psi_4(E_{i,j,k,l}) - \wh\eta_{4,J}\}^2$ for the studentization of $\wh\eta_{4,J}$. In light of Theorem B.1, for significance level $\alpha \in (0, 1)$, we can test $H_0^{(4)}: \eta_4=0$ versus $H_a^{(4)}:\eta_4\neq0$ as below:
\begin{equation}\label{test_eta4_test_decomp}
    \text{Reject~} H_0^{(4)} \text{~if~~} \big|n^{\lambda/2}~ \wh\eta_{4,J}/\wh\sigma_{4,J}\big| > \Phi^{-1}(1-\alpha/2). 
\end{equation}
Similar to the same-sender effects $\eta_3$, the finite-sample error control results for our test \eqref{test_eta4_test_decomp} are established, which, due to page limit, defer to Theorem B.2 with other discussions on test  \eqref{test_eta4_test_decomp} in the Supplementary Section B.

\section{Inference for sender-receiver effect \texorpdfstring{$\eta_5$}{eta5}}
\label{subsec::moment-based-test-of-network-effects::test-sender-receiver-effect} 
Consider a natural estimator of $\eta_5$:
\begin{align}
    \wh\eta_{5,n} :=&~ {n \choose 3}^{-1} \sum_{1 \leq i<j<k \leq n} h_5(E_{i,j,k}) - \Big\{{n \choose 2}^{-1} \sum_{1 \leq i<j \leq n} h_1(E_{i,j}) \Big\}^2,
    \notag
\end{align}
where $h_5(E_{i,j,k}) = \sum\nolimits^{\{i,j,k\}}_{\{i_1,i_2,i_3\}} e_{i_1,i_2}e_{i_2,i_3} / 6$.
Testing the sender-receiver effect $\eta_5$ significantly differs from those for $\eta_3$ and $\eta_4$. The key distinction is that under the null hypotheses, both $\hat\eta_{3,n}$ and $\hat\eta_{4,n}$ are degenerate, while the degeneracy status of $\hat\eta_{5,n}$ is even more complex. Specifically, it could be either non-degenerate or degenerate, contingent on the unknown underlying network. This necessitates different tests tailored to specific degeneracy status, making it essential to identify the degeneracy status beforehand. Therefore, we first develop a diagnostic test to determine the degeneracy status.

Similar to the analysis in Section \ref{subsubsection::Nontrivial-degeneracy-of-eta3}, we can define an $H_{5,n}$ by tweaking \eqref{H3_def}, replacing $h_3(\cdot)$ by $h_5(\cdot)$. Then, by Lemma A.2 in the Supplementary Materials, we have $\wh\eta_{5,n} - H_{5,n}=\tilde O_p(n^{-1}\log n)$. Define intermediate shorthand $g_{5,1}(x_i) := \Expect\{h_5(\tilde E_{i,j,k})|X_i=x_i\} - \Expect(e_{i,j}e_{j,k})$ and
$g_{5,2}(x_i,x_j) := \Expect\{h_5(\tilde E_{i,j,k})|X_i=x_i,X_j=x_j\} - g_{5,1}(x_i) - g_{5,1}(x_j) - \Expect(e_{i,j}e_{j,k})$.
We have
\begin{align}
    H_{5,n} - \eta_5
    =&~
    \frac{1}{n}\sum_{1 \leq i \leq n} g_{\eta_5,1}(X_i)
    +  {n \choose 2}^{-1}\sum_{1 \leq i<j \leq n}g_{\eta_5,2}(X_i,X_j)
    + 4n^{-1}\Expect\{g^2_{1,1}(X_1)\} + R_{H_5,n},
    \label{H_5n}
\end{align}
where 
\begin{align}
    g_{\eta_5,1}(X_i) 
    :=&~ 
    3g_{5,1}(X_i) - 4\mu_e g_{1,1}(X_i),
    \label{temp-def::g-eta5-1}
\end{align}
and
$g_{\eta_5,2}(X_i,X_j) := 3g_{5,2}(X_i,X_j) - 2\mu_e g_{1,2}(X_i,X_j) - 4n^{-1}(n-1)g_{1,1}(X_i)g_{1,1}(X_j)$.
In \eqref{H_5n}, the remainder $R_{H_5,n} = \tilde O_p(n^{-3/2}\log^{3/2}n)$ as shown by Lemma A.3 in the Supplementary Materials. 
    
To characterize the degeneracy status of $\hat\eta_{5,n}$, we first define 
\begin{align}
    \xi_{5,1} := \{\var(g_{\eta_5,1}(X_1))\}^{1/2}.
    \label{equation::xi51-def}
\end{align}
Now, we underscore the key distinction in the inference for $\eta_5$: unlike the cases for $\eta_3$ and $\eta_4$, the null hypothesis $H_0^{(5)}: \eta_5=0$ does \emph{not} imply the degeneracy of the natural point-estimation $\hat\eta_{5,n}$. Specifically, in \eqref{temp-def::g-eta5-1}, we have
\begin{align*}
    3g_{5,1}(X_i) 
    = &~
    \Expect\{f(X_i,X_j)f(X_j,X_k)|X_i\} + \Expect\{f(X_j,X_i)f(X_k,X_j)|X_i\} 
    \notag\\
    &+ \Expect(f(X_i,X_j)|X_i)\Expect(f(X_j,X_i)|X_i)- 3\mu_e^2
\end{align*}
and 
\begin{equation}    
    \label{equation::g11-def} 
    4\mu_e g_{1,1}(X_i)=2\mu_e \Expect(f(X_i,X_j) + f(X_j,X_i)|X_i) - 4\mu_e^2
\end{equation} 
for distinct indices $i,j,k$, yet $3g_{5,1}(X_i) - 4\mu_e g_{1,1}(X_i)$ does not admit any further reductions, rendering the determinate status of $g_{\eta_5,1}(X_i)$ unattainable under $H_0^{(5)}$. Therefore, it is imperative to meticulously assess the degeneracy status based on the available data and devise different testing procedures for different scenarios. In fact, similar to Example \ref{example::eta3-indeterminant-degen-g2}, the following example elucidates the indeterminate degeneracy of $\hat\eta_{5,n}$ under $H_0^{(5)}$.

\begin{example}[Indeterminate degeneracy of $\hat\eta_{5,n}$ under $H_0^{(5)}$]
\label{example::eta5-indeterminant-degen-g1}
    Inherit notation from Theorem \ref{exch_2}.
    Additionally, set $\{\epsilon_{i,j}\}_{1 \leq \{i,j\} \leq n}$ to be i.i.d., mean-zero random variables, independent of everything else. Then,
    \begin{enumerate}[(i)]
        \item if we model $e_{i,j}=X_i + X_{(i,j)}+\epsilon_{i,j}$, then $\xi_{5,1} \equiv 0$; 
        \item if we model $e_{i,j}=X_i (X_j-1/2) +\epsilon_{i,j}$, then $\xi_{5,1} \geq\mathrm{Constant}>0$. 
    \end{enumerate}
\end{example}

Readers may raise two natural questions. First, in practice, how do we know which degeneracy status we are in? Second, how to properly design a test for each status, respectively? We address the first question in Section \ref{subsubsec::moment-based-test-of-network-effects::test-sender-receiver-effect::test-degeneracy} and the second question in Sections \ref{subsubsec::moment-based-test-of-network-effects::test-sender-receiver-effect::non-degenerate-case} and \ref{subsubsec::moment-based-test-of-network-effects::test-sender-receiver-effect::degenerate-case}.

\subsection{Diagnosis of degeneracy}
\label{subsubsec::moment-based-test-of-network-effects::test-sender-receiver-effect::test-degeneracy}

Following Section \ref{subsubsection::Nontrivial-degeneracy-of-eta3}, we define the \textit{degenerate} and \emph{non-degenerate} cases of $\hat\eta_{5,n}$ by $\xi_{5,1} \equiv 0$ and $\xi_{5,1} \geq {\rm Constant}>0$, respectively. To decide between these two statuses, we test hypothesis $H_0 : \xi_{5,1} = 0$ versus $H_a : \xi_{5,1} \geq {\rm Constant}>0$. Based on \eqref{temp-def::g-eta5-1}, we construct the estimator of $\xi_{5,1}^2$ as
\begin{align}
    \wh\xi^2_{5,1}
    :=&~
    n^{-1}\sum^n_{i=1}\Big\{3\wh g_{5,1}(X_i) - 4{n \choose 2}^{-1} \sum_{1 \leq k<l \leq n} h_1(E_{k,l})\wh g_{1,1}(X_i)\Big\}^2,
    \label{empirical-xi-5-1}
\end{align}
where $\wh g_{5,1}(X_i)={n-1 \choose 2}^{-1} \sum_{\substack{1 \leq j<k \leq n;j,k\neq i }} h_5(E_{i,j,k})-{n \choose 3}^{-1} \sum_{1 \leq l<r<s \leq n} h_5(E_{l,r,s})$ and 
\begin{align}
    \wh g_{1,1}(X_i)=(n-1)^{-1} \sum_{\substack{1 \leq j \leq n;j\neq i }} h_1(E_{i,j})-{n \choose 2}^{-1} \sum_{1 \leq k<l \leq n} h_1(E_{k,l}).
    \label{empirical-g11}
\end{align}
We first establish the concentration result of $\wh\xi^2_{5,1}$ as follows.
\begin{proposition}
    \label{thm_var5_concentration} Consider an exchangeable network satisfying Definition \ref{def::WEdef} which admits the universal representation in Theorem \ref{exch_2}. Under Assumption \ref{assumption::sub-exp}, we have $\wh\xi^2_{5,1} -\xi_{5,1}^2=\tilde O_p(n^{-1/2}\log^{1/2} n)$.
\end{proposition}
For any specified $\alpha \in (0, 1)$, Proposition \ref{thm_var5_concentration} enables the testing of $H_0 : \xi_{5,1} = 0$ against $H_a : \xi_{5,1} \geq {\rm Constant} > 0$ by employing the following decision rule. 
\begin{equation}
    \text{Reject~} H_0 \text{~if~~} \wh\xi^2_{5,1} > C n^{-1/2}\log^{1/2} n,
    \label{eta5_test_degen}
\end{equation}
for some prespecified constant $C>0$. Proposition \ref{thm_var5_concentration} ensures that the type-I error rate of test \eqref{eta5_test_degen} is $O(1/n)$ and the type-II error rate of the test is $o(1)$ as $n$ increases.

\subsection{The non-degenerate case}
\label{subsubsec::moment-based-test-of-network-effects::test-sender-receiver-effect::non-degenerate-case}
    
Under the non-degenerate case, we have the following result regarding the asymptotic normality of $\wh\eta_{5,n}$.
\begin{theorem}
\label{thm_eta5_AN} 
    Under the conditions of Proposition \ref{thm_var5_concentration}, when $\xi_{5,1} \geq {\rm Constant}>0$, $\wh\eta_{5,n}$ is asymptotically normal:
    \begin{equation*}
        \sqrt{n}(\wh\eta_{5,n}-\eta_5) \xrightarrow{d} N(0, \xi_{5,1}^2).
    \end{equation*}
\end{theorem}
We can use $\wh\xi^2_{5,1}$ in \eqref{empirical-xi-5-1} as a natural variance estimator for $\wh\eta_{5,n}$, which can then be used in the subsequent studentization procedure. For the studentized $\wh\eta_{5,n}$, we have the following Berry-Esseen type bound. 
\begin{theorem}
\label{thm_eta5_BE_nondegen} 
    Under the conditions of Proposition \ref{thm_var5_concentration}, when $\xi_{5,1} \geq {\rm Constant}>0$, for some constant $C>0$, we have
    $$\sup_x\big|\Prob\{\sqrt{n}(\wh\eta_{5,n}-\eta_5)/\wh\xi_{5,1}\leq x\}-\Phi(x)\big| \leq Cn^{-1/2}\log n.$$
\end{theorem}
For any $\alpha \in (0, 1)$, Theorem \ref{thm_eta5_BE_nondegen} immediately leads to the test on the hypothesis in \eqref{nulls} as:
\begin{equation}
    \text{Reject~} H_0^{(5)} \text{~if~~} \big|\sqrt{n}\wh\eta_{5,n}/\wh\xi_{5,1}\big| > \Phi^{-1}(1-\alpha/2).
    \label{test_eta5_test}
\end{equation}
Utilizing Theorem \ref{thm_eta5_BE_nondegen}, the finite-sample error controls for our test \eqref{test_eta5_test} are established below. 
\begin{theorem}
\label{thm_var5_concen} 
    Under the conditions of Theorem \ref{thm_eta5_BE_nondegen}, applying test \eqref{test_eta5_test}, we have: 
    \begin{enumerate}[(i)]
    \item The probability of rejecting $H_0^{(5)}$ when it is true is $\alpha + O(n^{-1/2}\log n)$.
    \item The probability of failing to reject $H_0^{(5)}$ under $H_a^{(5)}$ is $o(1)$ when $\eta_5=\omega(n^{-1/2})$.
    \end{enumerate}
\end{theorem}
The finite-sample type-I error rate control in Theorem \ref{thm_var5_concen} coincides with the result in Theorem \ref{thm_eta3_imcp_concen} and B.2 (with $\lambda=1$). Theorem \ref{thm_var5_concen} not only implies that test \eqref{test_eta5_test} controls the type-I error rate with respect to the nominal level, but also confirms its power consistency. Moreover, it can effectively distinguish between $H_0^{(5)}:\eta_5=0$ and $H_a:\eta_5=\omega(n^{-1/2})$ in the non-degenerate case.

\subsection{The degenerate case}
\label{subsubsec::moment-based-test-of-network-effects::test-sender-receiver-effect::degenerate-case}

Similar to $\hat\eta_{3,n}$ and $\hat\eta_{4,n}$, the degree of degeneracy associated with $\hat\eta_{5,n}$ may also exceed $1$. Due to page limit, we relegate concrete examples to Section D.2 in the Supplementary Materials. Following the method proposed in Section \ref{subsubsection::Testing-using-the-U-statistics-reduction}, we consider a testing procedure based on reduced network moments. The reduced test statistic before studentization is defined as $\wh\eta_{5,J} := n^{-\lambda} \sum_{(i,j,k,l) \in J_{n,\lambda}} \psi_5(E_{i,j,k,l}),$ where $J_{n,\lambda}$ is defined in \eqref{incmp_3}, and $\psi_5(E_{i,j,k,l})$ is defined analogously to $\psi_3(E_{i,j,k,l})$ in \eqref{equation::eta_3n_reform} by replacing $h_3(E_{i_1,i_2,i_3})$ with $h_5(E_{i_1,i_2,i_3})$. Analogous to \eqref{incmp_3_decmp}, here, we decompose $\wh\eta_{5,J}$ as
\begin{equation}
    \wh\eta_{5,J} - \eta_5 = (\wh\eta_{5,n} - \eta_5) + n^{-\lambda} \sum_{(i,j,k,l) \in J_{n,\lambda}} \{\psi_5(E_{i,j,k,l})-\wh\eta_{5,n}\}.
\label{incmp_5_decmp}
\end{equation}
Likewise, we will show the second term on the right-hand side of \eqref{incmp_5_decmp} is dominating. Define the variance estimator for studentization as $\wh\sigma^2_{5,J}=n^{-\lambda} \sum_{(i,j,k,l) \in J_{n,\lambda}}\{ \psi_5(E_{i,j,k,l}) - \wh\eta_{5,J}\}^2$. We derive the Berry-Esseen type bound for the studentization in Theorem \ref{thm_eta5_imcp_concen_split-1}.
\begin{theorem}
\label{thm_eta5_imcp_concen_split-1} 
    Under the conditions of Proposition \ref{thm_var5_concentration}, when $\xi_{5,1} \equiv 0$, for $\lambda \in [1,2)$ such that $n^\lambda \in \mathbb{Z}$, we have 
    $$\sup_x\Big|\Prob\big[n^{\lambda/2}(\wh\eta_{5,J}-\eta_5)/\wh\sigma_{5,J}\leq x\big]-\Phi(x)\Big| \leq C(n^{\lambda/2-1}+n^{-\lambda/2})\log n,$$
    for some constant $C>0$.
\end{theorem}
Based on Theorem \ref{thm_eta5_imcp_concen_split-1}, we propose the following procedure for testing $H_0^{(5)}: \eta_5=0$ under the degenerate case. For any $\alpha \in (0, 1)$, we 
\begin{equation}
    \text{Reject~} H_0^{(5)} \text{~if~~} |n^{\lambda/2}~ \wh\eta_{5,J}/\wh\sigma_{5,J}\big| > \Phi^{-1}(1-\alpha/2).
    \label{test_eta5_test_decomp}
\end{equation}
Next, we have the finite-sample error control results for our test \eqref{test_eta5_test_decomp}.
\begin{theorem}
\label{thm_eta5_imcp_concen} 
    Under the conditions of Theorem \ref{thm_eta5_imcp_concen_split-1}, test \eqref{test_eta5_test_decomp} admits:
    \begin{enumerate}[(i)]
        \item the probability of rejecting $H_0^{(5)}$ when it is true is $\alpha + O((n^{\lambda/2-1}+n^{-\lambda/2})\log n)$.
        \item the probability of failing to reject $H_0^{(5)}$ under $H_a^{(5)}$ is $o(1)$ when $\eta_5=\omega(n^{-\lambda/2})$.
    \end{enumerate}
\end{theorem}

Similar to Theorem \ref{thm_eta3_imcp_concen}, Theorem \ref{thm_eta5_imcp_concen} characterizes the trade-off between the accuracy in controlling the type-I error rate and the power of the proposed test \eqref{test_eta5_test_decomp}. 
While (ii) recommends setting $\lambda$ near $2$ to enhance the power of our test \eqref{test_eta5_test_decomp}, for smaller networks, however, (i) advises setting $\lambda$ closer to $1$ to ensure more accurate type-I error rate control, where $\Prob_{H_0^{(5)}}\{\text{Reject~} H_0^{(5)}\} = \alpha + O(n^{-1/2}\log n)$ when $\lambda=1$.

\section{Inference for reciprocity effect \texorpdfstring{$\eta_2$}{eta2}}
\label{subsec::moment-based-test-of-network-effects::test-reciprocity-effect}

We study the reciprocity effect for last, primarily due to the intricacy arising from the dependency between $e_{i,j}$ and $e_{j,i}$, even when conditioned on $\{X_i\}_{1 \leq i \leq n}$. The point estimator of $\eta_2$ takes the following form
\begin{equation}
    \wh\eta_{2,n} := {n \choose 2}^{-1} \sum_{1 \leq i<j \leq n} h_2(E_{i,j}) - \Big\{{n \choose 2}^{-1} \sum_{1 \leq i<j \leq n} h_1(E_{i,j}) \Big\}^2,
    \label{eta2-estimator}
\end{equation}
where $h_2(E_{i,j}) = e_{i,j} e_{j,i}$. We face two technical issues. First, akin to $\hat\eta_{5,n}$, the estimator $\hat\eta_{2,n}$ also admits indeterminate degeneracy under the null hypothesis. Second, owing to the constraint $\{X_{(i,j)} = X_{(j,i)}\}_{1\leq i<j\leq n}$ in \eqref{eqn::Aldous-Hoover-representation}, $\rho_{i,j}$ and $\rho_{j,i}$ remain dependent even after conditioning on $\{X_i\}_{1 \leq i \leq n}$.
This is very different from, for instance, $\rho_{i,j}$ and $\rho_{j,k}$ in the inference for $\eta_5$. As a result, the dependency structure here is considerably more complex. Thus, we introduce an additional regularity condition concerning the distribution tail of $\rho_{i,j}\rho_{j,i}$.

\begin{assumption} 
\label{assumption::sub-exp-for-eta2}
    Recall that $\rho_{i,j}= e_{i,j}-\Expect(e_{i,j}|X_i,X_j)$ for each pair $(i,j)$. Conditioning on $\{X_i\}_{1 \leq i \leq n}$, we assume that $\{\rho_{i,j}\rho_{j,i}\}_{1 \leq i<j \leq n}$ are sub-exponential random variables.
\end{assumption}

Assumption \ref{assumption::sub-exp-for-eta2} is a very mild regularity condition concerning the tail behavior of the discrepancy between $e_{i,j}$ and its projection onto the space of $\{X_i\}_{1 \leq i \leq n}$. This assumption is found to be valid in the widely employed graphon models \citep{gao2016optimal}. It is also satisfied by the SRM \citep{Warner1979roundRobin,snijders1999srm} and conditionally independent dyad models \citep{chandrasekhar2016cid,graham2020cid}. 
    
Define $H_{2,n}$ by substituting $E_{i,j}$ in \eqref{eta2-estimator} with $\tilde E_{i,j}$. By Lemma A.4 in the Supplementary Materials, we have $\wh\eta_{2,n} - H_{2,n}=\Expect(\rho_{i,j} \rho_{j,i}) + \tilde O_p(n^{-1}\log n)$. Following Assumption \ref{assumption::sub-exp-for-eta2}, we obtain 
$$H_{2,n} + \Expect(\rho_{i,j} \rho_{j,i}) - \eta_2  =
\frac{1}{n}\sum_{1 \leq i \leq n} g_{1,\eta_2}(X_i)+ 
{n \choose 2}^{-1}\sum_{1 \leq i<j \leq n}g_{\eta_2,2}(X_i,X_j) + R_{H_2,n},$$
where 
\begin{align}
    g_{\eta_2,1}(X_i) := 2g_{2,1}(X_i) - 4\mu_e g_{1,1}(X_i)
    \label{equation::geta21-def}
\end{align}
and $g_{\eta_2,2}(X_i,X_j) := 2g_{2,2}(X_i,X_j) - 2\mu_e g_{1,2}(X_i,X_j) - 4n^{-1}(n-1)g_{1,1}(X_i)g_{1,1}(X_j)$. Here, we employ the shorthand $g_{2,2}(x_i,x_j) := \Expect\{h_2(\tilde E_{i,j})|X_i=x_i,X_j=x_j\} - g_{2,1}(x_i) - g_{2,1}(x_j) - \Expect(e_{i,j}e_{j,k} - \rho_{i,j} \rho_{j,i})$ and $g_{2,1}(x_i) := \Expect\{h_2(\tilde E_{i,j})|X_i=x_i\} - \Expect(e_{i,j}e_{j,i} - \rho_{i,j} \rho_{j,i})$. By Lemma A.4 in the Supplementary Materials, the remainder $R_{H_2,n} = \tilde O_p(n^{-1}\log n)$. To characterize the degeneracy status of $\hat\eta_{2,n}$, we define
\begin{align}
    \xi_{2,1} := \{\var(g_{\eta_2,1}(X_1))\}^{1/2}.
    \label{equation::xi21-def}
\end{align}
Echoing the observation in Section \ref{subsec::moment-based-test-of-network-effects::test-sender-receiver-effect}, the condition $\eta_2=0$ does not conclusively determine whether $\xi_{2,1}$ equals zero. Therefore, under the null hypothesis $H_0^{(2)}$, $\hat\eta_{2,n}$ exhibits the same indeterminate degeneracy, as exemplified in Example \ref{example::eta2-indeterminant-degen-g1}, which is similar to Example \ref{example::eta5-indeterminant-degen-g1}.

\begin{example}[Indeterminate degeneracy of $\hat\eta_{2,n}$ under $H_0^{(2)}$]
\label{example::eta2-indeterminant-degen-g1}
    Inherit the notation from Theorem \ref{exch_2}.
    Additionally, set $\{\epsilon_{i,j}\}_{1 \leq \{i,j\} \leq n}$ to be i.i.d., mean-zero random variables, independent of everything else.  
    Then
    \begin{enumerate}[(i)]
        \item if we model $e_{i,j}=X_{(i,j)}+\epsilon_{i,j}$, then $\xi_{2,1} \equiv 0$; 
        \item if we model $e_{i,j}=X_i-X_j+\sqrt{2}X_{(i,j)}+\epsilon_{i,j}$, then $\xi_{2,1} \geq \rm Constant > 0$. 
    \end{enumerate}
\end{example}
Since the distribution of $\hat\eta_{2,n}$ is contingent on its degeneracy status, we propose a diagnostic test on degeneracy for selecting the proper downstream inference procedure.

\subsection{Diagnosis of degeneracy} 
\label{subsubsec::moment-based-test-of-network-effects::test-reciprocity-effect::test-degeneracy}
    
Define the \textit{degenerate} and \emph{non-degenerate} cases of $\hat\eta_{2,n}$ by $\xi_{2,1} \equiv 0$ and $\xi_{2,1} \geq {\rm Constant}>0$, respectively. We propose the following diagnostic test for $H_0 : \xi_{2,1} = 0$ versus $H_a : \xi_{2,1} \geq {\rm Constant}>0$.
\begin{equation}
    \text{Reject~} H_0 \text{~if~~} \wh\xi^2_{2,1} > C n^{-1/2}\log^{1/2} n,
    \label{eta2_test_degen}
\end{equation}
where $\wh\xi^2_{2,1}=n^{-1}\sum^n_{i=1}\{2\wh g_{2,1}(X_i) - 4{n \choose 2}^{-1} \sum_{1 \leq k<l \leq n} h_1(E_{k,l})\wh g_{1,1}(X_i)\}^2,$ with $\wh g_{2,1}(X_i)$ defined by replacing $h_1(E_{i,j})$ with $h_2(E_{i,j})$ in \eqref{empirical-g11}; and the positive constant $C$ is prespecified. By Proposition \ref{thm_var2_concentration} below, the type-I error rate of test \eqref{eta2_test_degen} is $O(1/n)$ and the type-II error rate of the test is $o(1)$ as $n$ increases.
\begin{proposition}
    \label{thm_var2_concentration} 
    Consider an exchangeable network satisfying Definition \ref{def::WEdef} which admits the universal representation in Theorem \ref{exch_2}. Under Assumption \ref{assumption::sub-exp} and Assumption \ref{assumption::sub-exp-for-eta2}, we have $\wh\xi^2_{2,1}-\xi_{2,1}^2=\tilde O_p(n^{-1/2}\log^{1/2} n)$.
\end{proposition}

\subsection{The non-degenerate case}
\label{subsubsec::moment-based-test-of-network-effects::test-reciprocity-effect::non-degenerate-case}

Following the approach outlined in Section \ref{subsubsec::moment-based-test-of-network-effects::test-sender-receiver-effect::non-degenerate-case}, we first establish the asymptotic normality of $\wh\eta_{2,n}$ in Theorem \ref{thm_eta2_AN}. Then, we derive a Berry-Esseen type bound for the studentized estimator in Theorem \ref{thm_var2_concen_split-1}, and proceed to develop the testing procedure. Lastly, we elucidate the finite-sample error controls of our test in Theorem \ref{thm_var2_concen}. This not only ensures accurate control over the type-I error rate but also affirms the power consistency of our test on the reciprocity effect under the non-degenerate case.
\begin{theorem}
\label{thm_eta2_AN} 
    Under the conditions of Proposition \ref{thm_var2_concentration}, when $\xi_{2,1} \geq {\rm Constant}>0$, $\wh\eta_{2,n}$ is asymptotically normal:
    \begin{equation*}
    \sqrt{n}(\wh\eta_{2,n}-\eta_2) \xrightarrow{d} N(0, \xi^2_{2,1}).
    \end{equation*}
\end{theorem}
    
\begin{theorem}
\label{thm_var2_concen_split-1} 
    Under the conditions of Proposition \ref{thm_var2_concentration}, when $\xi_{2,1} \geq {\rm Constant}>0$, for some constant $C>0$, we have $$\sup_x\big|\Prob\{\sqrt{n}(\wh\eta_{2,n}-\eta_2)/\wh\xi_{2,1}\leq x\}-\Phi(x)\big| \leq Cn^{-1/2}\log n.$$
\end{theorem}

Theorems \ref{thm_eta2_AN} and \ref{thm_var2_concen_split-1} facilitate testing for $H_0^{(2)}: \eta_2=0$ versus $H_a^{(2)}:\eta_2\neq0$ as follows. For any given $\alpha \in (0, 1)$, \begin{equation} \label{test_eta2_test} \text{Reject~} H_0^{(2)} \text{~if~~} \big|\sqrt{n}\wh\eta_{2,n}/\wh\xi_{2,1}\big| > \Phi^{-1}(1-\alpha/2), 
\end{equation}
whose finite-sample error controls are established in the following theorem. 
\begin{theorem} 
\label{thm_var2_concen}
    Under the conditions of Theorem \ref{thm_var2_concen_split-1}, our test \eqref{test_eta2_test} admits: 
    \begin{enumerate}[(i)]
    \item The probability of rejecting $H_0^{(2)}$ when it is true is $\alpha + O(n^{-1/2}\log n)$.
    \item The probability of failing to reject $H_0^{(2)}$ under $H_a^{(2)}$ is $o(1)$ when $\eta_2=\omega(n^{-1/2})$.
    \end{enumerate}
\end{theorem}

\subsection{The degenerate case}
\label{subsubsec::moment-based-test-of-network-effects::test-reciprocity-effect::degenerate-case}

Similar to Section \ref{subsubsec::moment-based-test-of-network-effects::test-sender-receiver-effect::degenerate-case}, the degeneracy order of $\hat\eta_{2,n}$ may exceed degree $1$, which is confirmed by examples in the Supplementary Section D.2. Likewise, we consider the reduced test statistic before studentization defined as $\wh\eta_{2,J} := n^{-\lambda} \sum_{(i,j,k,l) \in J_{n,\lambda}} \psi_2(E_{i,j,k,l})$, and variance estimator  $\wh\sigma^2_{2,J}=n^{-\lambda} \sum_{(i,j,k,l) \in J_{n,\lambda}}\{ \psi_2(E_{i,j,k,l}) - \wh\eta_{2,J}\}^2$, where $\psi_2(E_{i,j,k,l}) = \sum\nolimits_{\{i_1,i_2\}\in\{i,j,k,l\}}h_2(E_{i_1,i_2})/3 - h_6(E_{i,j,k,l}) + r(E_{i,j,k,l})$ with $r(E_{i,j,k,l})$ as defined in Section \ref{subsubsection::Testing-using-the-U-statistics-reduction}. Similarly to Theorem \ref{thm_eta5_imcp_concen_split-1}, Theorem  \ref{thm_eta2_imcp_concen_split-1} establishes Berry-Esseen type bound for appropriately studentized $\wh\eta_{2,J}$.

\begin{theorem} 
\label{thm_eta2_imcp_concen_split-1} 
    Under the conditions of Proposition \ref{thm_var2_concentration}, when $\xi_{2,1} = 0$, for $\lambda \in [1,2)$ such that $n^\lambda \in \mathbb{Z}$, we have $$\sup_x\Big|\Prob\big[n^{\lambda/2}(\wh\eta_{2,J}-\eta_2)/\wh\sigma_{2,J}\leq x\big]-\Phi(x)\Big| \leq C(n^{\lambda/2-1}+n^{-\lambda/2})\log n,$$ for some constant $C>0$.
\end{theorem}
    
For any specified $\alpha \in (0, 1)$, Theorem \ref{thm_eta2_imcp_concen_split-1} provides a test for $ H_0^{(2)}$ as follows: \begin{equation}  \text{Reject~} H_0^{(2)} \text{~if~~} |n^{\lambda/2}~ \wh\eta_{2,J}/\wh\sigma_{2,J}\big| > \Phi^{-1}(1-\alpha/2). \label{test_eta2_test_decomp}
\end{equation}  

Building on Theorem \ref{thm_eta2_imcp_concen_split-1}, we derive finite-sample error controls for our test \eqref{test_eta2_test_decomp}, paralleling Theorems \ref{thm_eta3_imcp_concen} and \ref{thm_eta5_imcp_concen}.
It also suggests that setting a smaller value of $\lambda$ enhances the type-I error control accuracy, while increases the minimum separation condition required for power consistency.

\begin{theorem}
\label{thm_eta2_imcp_concen}   
    Under the conditions of Theorem \ref{thm_eta2_imcp_concen_split-1}, for our test \eqref{test_eta2_test_decomp}, we have:
    \begin{enumerate}[(i)]
    \item The probability of rejecting $H_0^{(2)}$ when it is true is $\alpha + O((n^{\lambda/2-1}+n^{-\lambda/2})\log n)$.
    \item The probability of failing to reject $H_0^{(2)}$ under $H_a^{(2)}$ is $o(1)$ when $\eta_2=\omega(n^{-\lambda/2})$.
 \end{enumerate}
\end{theorem}

\section{Optimality of testing network effects}
\label{section::optimality}   

In this section, we present lower bound results on the minimal separation conditions in testing different network effects. To the best of our knowledge, our findings represent the first effort on establishing minimax optimality of consistent tests for network effects. By comparing the lower bound results with the finite-sample guarantees presented in earlier sections, we affirm the rate-optimality of our tests for various network effects.

\subsection{Optimality of testing same-sender effect $\eta_3$ and same-receiver effect $\eta_4$}
\label{subsection::optimality-for-testing-ets3and4}

We begin by developing the lower bounds on the minimal separation of $\eta_3$ or $\eta_4$ for distinguishing the network with no same-sender or same-receiver effect from that with certain nonzero same-sender or same-receiver effect. Formally, we have the following result.
        
\begin{theorem}    
\label{thm_eta3_lower_bound} 
    For any $\alpha \in (0,1)$ and $\ell\in \{3,4\}$, there exists exchangeable network models $f_0$ under $H_0^{(\ell)}$ and $f_a$ under $H_a^{(\ell)}$, satisfying $\eta_{\ell}=0$ under $f_0$ and $\eta_{\ell}=O(n^{-1})$ with $\xi_{\ell,1} = O(n^{-1/2})$ under $f_a$. Any test $\mathcal{T}$ with the type-I error rate not exceeding $\alpha$ admits $$\mathbb{P}(\textit{Reject~} H_0^{(\ell)}|H_0^{(\ell)})+\mathbb{P}(\textit{Fail to reject~} H_0^{(\ell)}|H_a^{(\ell)})\geq \mathrm{Constant}>0.$$
\end{theorem}

\begin{remark}
\label{remark::lower-bound-parameter-space-xi-1}
    Before any further discussion, it is important that we address a common question that may arise among readers regarding Theorem \ref{thm_eta3_lower_bound}. This theorem establishes a fundamental lower bound that constrains the performance of any test. At a first glance, $\xi_{\ell,1}$ appears to be a quantity specific to our method. Is this a conceptual error? In fact, observe that $\xi_{\ell,1}$ can be reformulated as a functional of the underlying model by \eqref{equation::xi31-def} for $\xi_{3,1}$ and (SB.2) for $\xi_{4,1}$. Hence, putting a condition on $\xi_{\ell,1}$ should be seen as imposing a constraint directly on the model itself, not exclusively through our method.
\end{remark}
    
The proof of Theorem \ref{thm_eta3_lower_bound} is given in Supplementary Section C. At a high-level, similar to the approaches in \cite{cai2013optimal} and \cite{shao2022higher}, we carefully constructed graphon models $f_0$ and $f_a$, incorporating different distributions of latent variables that generate exchangeable networks with and without the desired network effect. We then lower bound the performance of the optimal likelihood ratio test in distinguishing such $f_0$ against $f_a$. But here, we need a novel construction of the least favorable configuration completely different from that used to prove Theorem 3 in \cite{shao2022higher}. This new construction ensures that $\xi_{\ell,1} = O(n^{-1/2})$. In fact, Part (i) in both Theorems \ref{thm_eta3_imcp_concen} and B.2 remains valid if $\xi_{\ell,1}=O(n^{-1/2})$ for $\ell \in \{3,4\}$, as is evident from their proofs. Hence, the condition on $\xi_{\ell,1}$ in Theorem \ref{thm_eta3_lower_bound} matches the assumption in Theorem \ref{thm_eta3_lower_bound} and therefore confirms that our tests on both the same-sender and same-receiver effect are nearly rate-optimal in power when $\lambda \approx 2$. On the other hand, an accurate control of the type-I error rate near $\alpha$ prefers $\lambda=1$. It is important to note that choosing $\lambda<2$ reinstates normality for the incomplete U-statistics \citep{weber1981incomplete, Chen2019U, shao2023U-statistics}. This, however, comes at the expense of increased variance, and may diminish the power. In practice, it is advisable for users to select a $\lambda\in (1,2)$, erring on the side of $1$ for smaller sample sizes where the accurate control of the type-I error is critical. This is further confirmed numerically in Section \ref{sec:typeIerror}.

\subsection{Optimality of testing sender-receiver effect $\eta_5$ and reciprocity effect $\eta_2$}

Next, in Theorem \ref{thm_eta5_lower_bound}, we show the lower bound results for testing the reciprocity effect $\eta_2$ and the sender-receiver effect $\eta_5$, presented as the minimum separation condition under $H_a^{(2)}$ and $H_a^{(5)}$, respectively. As addressed in Sections \ref{subsec::moment-based-test-of-network-effects::test-sender-receiver-effect} and \ref{subsec::moment-based-test-of-network-effects::test-reciprocity-effect}, we encounter the indeterminate degeneracy of $\hat\eta_{\ell,n}$ under $H_0^{(\ell)}$, for $\ell\in \{2,5\}$. Like in Theorem \ref{thm_eta5_lower_bound}, here, we also observe different lower bound results in different degenerate and non-degenerate cases. 
    
\begin{theorem}
\label{thm_eta5_lower_bound} 
    For any $\alpha \in (0,1)$ and $\ell\in \{2,5\}$, we have
    \begin{enumerate}[(i)]
    \item when $\xi_{\ell,1} = O(n^{-1/2})$, there exists exchangeable networks $f_{0}$ under $H_0^{(\ell)}$ and $f_{a}$ under $H_a^{(\ell)}$, satisfying $\eta_\ell=0$ under $f_{0}$ and $\eta_\ell=O(n^{-1})$ under $f_{a}$, or
    \item when $\xi_{\ell,1} \geq\mathrm{Constant}>0$, there exists exchangeable networks $f_{0'}$ under $H_0^{(\ell)}$ and $f_{a'}$ under $H_a^{(\ell)}$, satisfying $\eta_\ell=0$ under $f_{0'}$ and $\eta_\ell=O(n^{-1/2})$ under $f_{a'}$,
    \end{enumerate}
    such that any test $\mathcal{T}$ with the type-I error rate not exceeding $\alpha$ admits
    $$\mathbb{P}(\textit{Reject~} H_0^{(\ell)}|H_0^{(\ell)})+\mathbb{P}(\textit{Fail to reject~} H_0^{(\ell)}|H_a^{(\ell)})\geq \mathrm{Constant}>0.$$
\end{theorem}
  
Combining Theorem \ref{thm_eta5_lower_bound} with the upper bound results in Theorems \ref{thm_var5_concen} and \ref{thm_var2_concen} shows that our method is rate-optimal under the non-degenerate case. In the proofs of Theorems \ref{thm_eta5_imcp_concen} and \ref{thm_eta2_imcp_concen}, we find that the finite-sample type-II error rate control of our method remains valid if we relax the condition $\xi_{\ell,1}=0$ to $\xi_{\ell,1}=O(n^{-1/2})$ for $\ell\in \{2,5\}$. This aligns with the condition in part (i) of Theorem \ref{thm_eta5_lower_bound} and confirms the near rate-optimality of our method for $\lambda$ close to $2$ under degeneracy. Recall that Remark \ref{remark::lower-bound-parameter-space-xi-1} pointed out an interesting fact that $\xi_{\ell,1}$, which indicates the degeneracy status of our test statistic, reflects the fundamental hardness of the problem not only for our method, but for any method, because $\xi_{\ell,1}$ can be viewed as a functional of the model, here \eqref{temp-def::g-eta5-1} and \eqref{equation::xi51-def} for $\xi_{5,1}$ and by \eqref{equation::geta21-def} and \eqref{equation::xi21-def} for $\xi_{2,1}$.

\section{Simulations}
\label{section::simulation}    
    
As mentioned in Section \ref{section::intro}, available tools on network effects inference are scarce in the literature, predominantly confined to studies of SRM that presume additive models with specific parametric distributional constraints. In this section, we compare the finite-sample performance of our method, called {\tt NET} (short for {\bf N}etwork {\bf E}ffect {\bf T}est), with that of the widely used methods under SRM assumption: the ANOVA approach, called {\tt SRM-A} \citep{lashley1997significance}, and the likelihood based method, called {\tt SRM-L}, \citep{nestler2020maximum}. We employed \texttt{R} packages ``\texttt{TripleR}'' for {\tt SRM-A} and ``\texttt{srm}'' for {\tt SRM-L}, respectively. 

Throughout this section, we omit the super-index in $H_0^{(\ell)}$ and $H_a^{(\ell)}$ where clarity permits. We experimented on $\eta_2$, $\eta_3$, and $\eta_5$, choosing $\eta_3$ over $\eta_4$ due to their similar testing procedures. Both ANOVA and likelihood-based methods primarily concentrate on individual SRM parameters, like $\sigma_{ab}$ and $\sigma^2_\epsilon$ in model \eqref{eqn::social-relations-model}, without providing readily applicable tests for the reciprocity effect $\eta_2$. Thus, we exclusively present results of our method for testing $\eta_2$. For the diagnosis of degeneracy in Sections \ref{subsubsec::moment-based-test-of-network-effects::test-sender-receiver-effect::test-degeneracy} and \ref{subsubsec::moment-based-test-of-network-effects::test-reciprocity-effect::test-degeneracy}, we set the prespecified constant $C=1$. We generated observable edges $\{e_{i,j}\}_{1\leq \{i,j\}\leq n}$ using i.i.d. $\{a_i\}_{1 \leq i \leq n}$, $\{b_i\}_{1 \leq i \leq n}$, $\{\gamma_{(i,j)}=\gamma_{(j,i)}\}_{1 \leq i<j \leq n}$, and errors $\{\epsilon_{i,j}\}_{1\leq \{i,j\}\leq n}$: in the ``normal configuration'', $a_i\sim N(1,1)$, $b_i\sim N(1,1)$, $\gamma_{(i,j)}\sim N(1,1)$, and $\epsilon_{i,j}\sim N(0,1)$, while in the ``Poisson configuration'',  $a_i$, $b_i$, $\gamma_{(i,j)}$ and $\epsilon_{i,j}$ all followed Poisson$(1)$. We considered both additive and multiplicative models as follows.
\begin{enumerate}[(a)] 
    \item \label{test_eta2_add} Testing $\eta_2$: set $e_{i,j}= a_i+b_j+c\gamma_{(i,j)}+\epsilon_{i,j}$. Under $H_0$, $c=0$ and under $H_a$, $c \in \{0.05,0.2,0.5,1,5\}^{1/2}$.
    \item \label{test_comp_a} Testing $\eta_3$: set $e_{i,j}=ca_i+\epsilon_{i,j}$.
    Under $H_0$, $c=0$ and under $H_a$, $c \in \{0.05,0.2,0.5,1,5\}^{1/2}$. 
    \item \label{test_comp_d} Testing $\eta_5$: set $e_{i,j}=c(a_i-d)(a_j-d)+\epsilon_{i,j}$. Under $H_0$, $c=d=1$ and under $H_a$, $d=0$ and $c \in \{0.05,0.2,0.5,1,5\}^{1/2}$.
\end{enumerate} 
Among them, Setting \eqref{test_comp_a} satisfies the additive model assumption of SRM. In contrast, Setting \eqref{test_comp_d} includes multiplicative components, thereby contravening SRM's assumptions. In our simulations, we set $\alpha=0.05$ and varied the network size with $n\in\{25, 50, 100, 200, 400\}$. Each experiment was repeated $1,000$ times for methods applicable to the specific settings. We assessed the numerical robustness of all methods by the ratio of producing non-NA output values in these $1,000$ numerical experiments. Additionally, whenever tests were constructed based on reduced network moments, we evaluated our method for distinct $\lambda \in \{1, 1.2, 1.6\}$.

\begin{figure}[h!]
    \centering
    \setlength{\abovecaptionskip}{-0.1cm}
    \includegraphics[width=\linewidth]{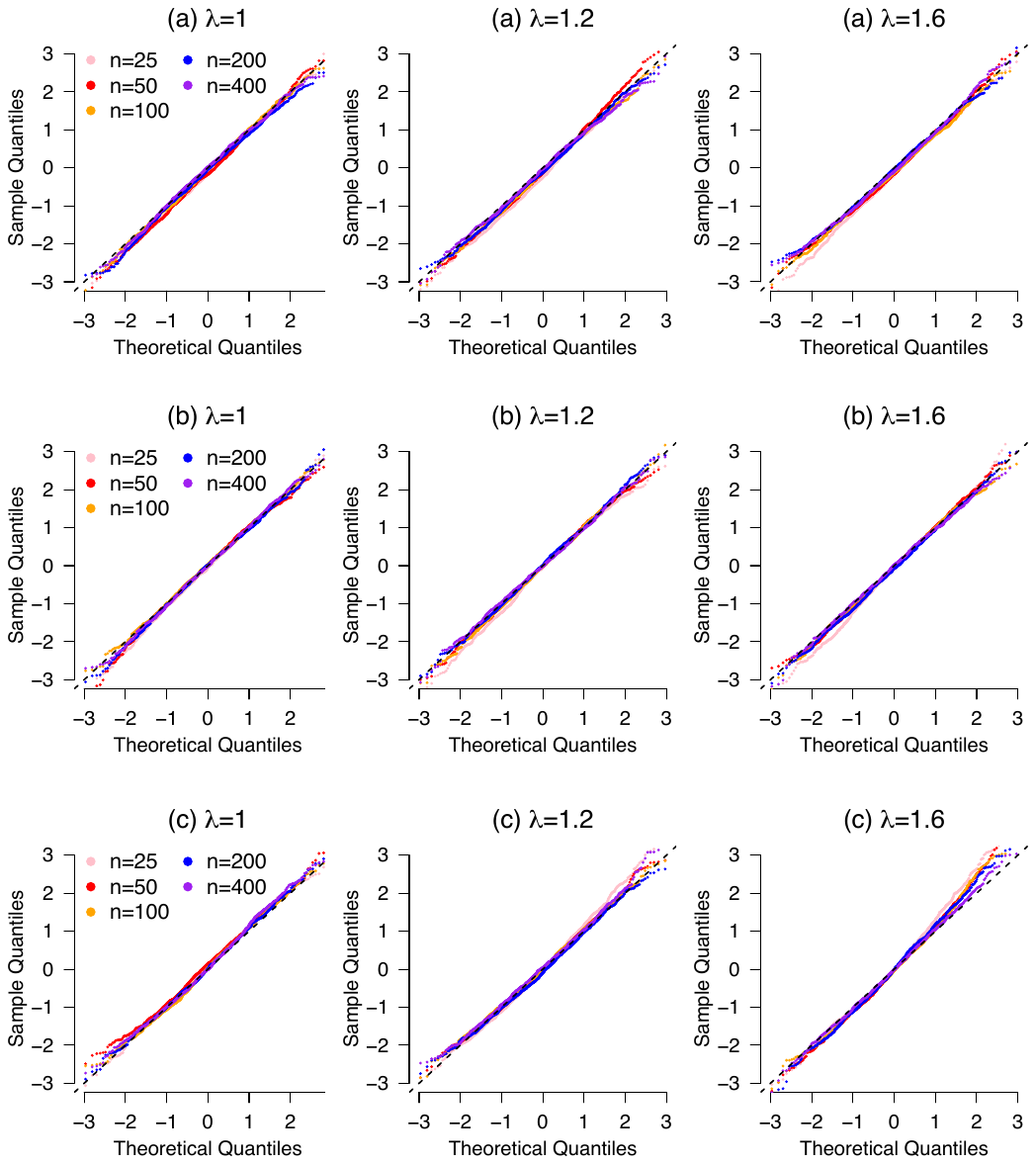} 
    \caption {Q-Q plots of the null distribution of test statistics under normal configurations across settings \eqref{test_eta2_add} to \eqref{test_comp_d}.} 
\label{fig::size-normal-setting-abcd} 
\end{figure}

Prior to evaluating the empirical type-I error rate and power, we first numerically examine our theoretical results on the normal approximation to the sampling distribution of proposed test statistics. As evidenced by the Q-Q plots in Figure \ref{fig::size-normal-setting-abcd}, the established approximations prove accurate across various network sizes and settings. In line with the finite sample results (e.g. Section \ref{Sec313}), setting $\lambda=1$ yields better approximation accuracy than higher values of $\lambda$, thereby enhancing the accuracy of type-I error rate control. We have deferred results for the Poisson configuration and further discussions on the practical choice of $\lambda$ to Section  E.1 in the Supplementary Materials.

\subsection{Type-I error rate}
\label{sec:typeIerror}

We now examine the empirical type-I error rates of all methods, as detailed in Table \ref{tab::smry_size_marginal_C=1_updated}. For testing $\eta_2$ in setting \eqref{test_eta2_add}, our {\tt NET} method effectively controls the type-I error rate close to the nominal level. In settings \eqref{test_comp_a} and \eqref{test_comp_d}, {\tt NET} with $\lambda=1$ outperforms other methods in accurately controlling the type-I error rate across all configurations, except for a minor inflation in small networks where $n=25$. Increasing $\lambda$ from $1$ to $1.6$ leads to a slight rise in the type-I error rate, but enhances test power, as further discussed in Sections \ref{sec:power} and E.1 in the Supplementary Materials. Under setting \eqref{test_comp_a}, the benchmark method {\tt SRM-A} tends to be conservative in small networks with $n=25$ or $50$, and less accurate in controlling the finite-sample type-I error rate in larger networks with $n=400$, compared to our method. The {\tt SRM-L} method struggles to control the type-I error rate in smaller networks with $n=25$ or $50$, as its validity depends on large sample approximations. Although its performance improves with larger $n$, the best existing \texttt{R} package for {\tt SRM-L} method requires excessive memory, preventing its application in networks with $n\geq 100$ nodes. 
  
\begin{table}[h!] \caption {Empirical type-I error rates of all methods at the nominal level $\alpha=0.05$. SRM-based methods are not suitable for testing $\eta_2$ and are thus indicated with a $\times$, along with experiments that exceed the memory limit.} \label{tab::smry_size_marginal_C=1_updated} 
    \begin{center}
    \begin{tabular}{cccccccccc}
    \toprule
    &&& \multicolumn{3}{c}{Normal configuration}                                &  & \multicolumn{3}{c}{Poisson configuration}      \\ \cmidrule{4-6} \cmidrule{8-10} 
    && LogTime & \eqref{test_eta2_add} &  $\eqref{test_comp_a}$ & $\eqref{test_comp_d}$ &  & \eqref{test_eta2_add} & $\eqref{test_comp_a}$ & $\eqref{test_comp_d}$ \\ \midrule
    $n=25$ & {\tt NET($\lambda=1$)} & -0.938 & 0.063 & 0.066  &   0.056    &  &  0.068 &  0.068   &  0.049    \\
    & {\tt NET($\lambda=1.2$)} & -0.486 & 0.068 & 0.059  &   0.076    &  &  0.061 & 0.066   &  0.068 \\
    & {\tt NET($\lambda=1.6$)} & 0.452 & 0.070 & 0.080  &   0.101    &  & 0.062 &  0.070   &  0.101 \\
    &{\tt SRM-A} & 0.574 & $\times$ & 0.005  &  0.325    &  &  $\times$ & 0.013    &  0.303    \\
    &{\tt SRM-L} & 7.459 & $\times$ & 0.278  &  0.634    &  & $\times$ &  0.270    &  0.607    \\
    \midrule
    $n=50$ & {\tt NET($\lambda=1$)} & -0.463 & 0.067 & 0.054   &  0.041    &  &  0.060 & 0.047   &  0.065    \\
    & {\tt NET($\lambda=1.2$)} & 0.094 & 0.070 & 0.052  &   0.046    &  & 0.060 &  0.061   &  0.062 \\
    & {\tt NET($\lambda=1.6$)} & 1.470 & 0.057 & 0.065  &   0.083    &  & 0.063 &  0.047   &  0.077 \\
    &{\tt SRM-A} & 1.809 & $\times$ & 0.027  &  0.528    &  & $\times$ &  0.026   &  0.562     \\
    &{\tt SRM-L} & 9.008 & $\times$ & 0.102  &  0.706   &  & $\times$ &  0.109    &  0.710    \\
    \midrule
    $n=100$  & {\tt NET($\lambda=1$)} & 0.003 &  0.048 & 0.050  &  0.047   &  &  0.055 & 0.039    &  0.038    \\
    & {\tt NET($\lambda=1.2$)} & 0.823 & 0.045 & 0.060  &   0.047    &  & 0.048 &  0.045   &  0.040 \\
    & {\tt NET($\lambda=1.6$)} & 2.549 & 0.051 & 0.047  &   0.083    &  &  0.065 & 0.050   &  0.061 \\
    &{\tt SRM-A} & 3.953 & $\times$ & 0.047  &   0.736    &  & $\times$ &  0.046   &  0.764     \\
    &{\tt SRM-L} & $\times$ & $\times$ & $\times$  &  $\times$   &  & $\times$ &  $\times$   &   $\times$    \\
    \midrule
    $n=200$ & {\tt NET($\lambda=1$)} & 0.631 &  0.053 & 0.051  &  0.059   &  & 0.055 &  0.051   &  0.044    \\
    & {\tt NET($\lambda=1.2$)} & 1.634 & 0.057 & 0.060  &   0.045    &  & 0.062 &  0.056   &  0.056 \\
    & {\tt NET($\lambda=1.6$)} & 3.651 & 0.035 & 0.055  &   0.077    &  &  0.051 & 0.054   &  0.057 \\
    &{\tt SRM-A} & 4.837 & $\times$ & 0.066  &  0.855   &  & $\times$ &  0.041    &  0.831    \\
    &{\tt SRM-L} & $\times$ & $\times$ & $\times$  &  $\times$   &  &  $\times$ & $\times$   &   $\times$    \\
    \midrule
    $n=400$ & {\tt NET($\lambda=1$)} & 1.343 &  0.054 & 0.058 &  0.055   &  & 0.061 &  0.050   &  0.044    \\
    & {\tt NET($\lambda=1.2$)} & 2.461 & 0.037 & 0.048  &   0.055    &  &  0.034 & 0.054   &  0.052 \\
    & {\tt NET($\lambda=1.6$)} & 4.762 & 0.053 & 0.049  &   0.054    &  &  0.057 & 0.049   &  0.057 \\
    &{\tt SRM-A} & 5.987 & $\times$ & 0.068  &  0.848   &  & $\times$ &  0.072   &  0.851    \\
    &{\tt SRM-L} & $\times$ & $\times$ &  $\times$  &  $\times$   &  &  $\times$ & $\times$   &  $\times$    \\
    \bottomrule
    \end{tabular}
    \end{center}
\end{table}

In setting \eqref{test_comp_d}, where the additive model assumption is violated, both {\tt SRM-A} and {\tt SRM-L} failed to control the type-I error rate, regardless of the network size $n$. This confirms the vulnerability of existing methods under model-misspecification. Moreover, the average empirical running times, as displayed in Table \ref{tab::smry_size_marginal_C=1_updated}, clearly illustrate the computational advantage of our method, even with larger choices of the tuning parameter $\lambda$. Furthermore, we observed that in over half of the Monte Carlo iterations, the existing package for {\tt SRM-A} encountered numerical difficulties, leading to {\tt NA} outputs, as elaborated in Table S1 in the Supplementary Materials. In contrast, our method exhibited a high degree of numerical robustness.

\subsection{Power}
\label{sec:power}
    
To evaluate the empirical power of our {\tt NET} method in comparison to benchmarks, we focus on networks with sizes $n\in\{50, 100\}$, because competing methods encounter computational challenges in larger networks, as discussed in Section \ref{sec:typeIerror}. In Table \ref{tab::smry_power_marginal_C=1_updated}, we present the empirical power of {\tt NET} under setting \eqref{test_eta2_add} and compare it with the empirical powers of other two methods under settings \eqref{test_comp_a}-\eqref{test_comp_d}. However, results for {\tt SRM-L} at $n=100$ are omitted as it exceeds the memory limit.

\begin{table}[h!] 
\caption{Empirical powers of the three methods. Setting \eqref{test_eta2_add} is only testable by {\tt NET}, with SRM-based methods marked with a $\times$. Setting \eqref{test_comp_a} satisfies the additive model assumption of SRM, whereas setting \eqref{test_comp_d} does not. Experiments exceeding memory limits are indicated with a $\times$. The background colors of power values correspond to their respective type-I error rate controls, detailed in Table \ref{tab::smry_size_marginal_C=1_updated}.} \label{tab::smry_power_marginal_C=1_updated} 
    \setlength\tabcolsep{4pt}
    \begin{center}
    \begin{tabular}{ccccccccccccc}
    \toprule
     && \multicolumn{5}{c}{$\eqref{test_eta2_add}$ - Normal configuration}          &  & \multicolumn{5}{c}{$\eqref{test_eta2_add}$ - Poisson configuration}      \\ \cmidrule{3-7} \cmidrule{9-13} 
    & $c^2$ & 0.05 & 0.2 & 0.5 & 1 & 5 &  & 0.05 & 0.2 & 0.5 & 1 & 5 \\ \midrule
    $n=50$ & {\tt NET($\lambda=1$)}  & \cellcolor{red!17} 0.063  & \cellcolor{red!17} 0.100  & \cellcolor{red!17} 0.397  & \cellcolor{red!17} 0.891 & \cellcolor{red!17} 1.000   &  &  \cellcolor{red!10} 0.053  & \cellcolor{red!10} 0.082 & \cellcolor{red!10} 0.385  & \cellcolor{red!10} 0.886 & \cellcolor{red!10} 1.000    \\
    & {\tt NET($\lambda=1.2$)}  & \cellcolor{red!20} 0.054  & \cellcolor{red!20} 0.092  & \cellcolor{red!20} 0.387  & \cellcolor{red!20} 0.891 & \cellcolor{red!20} 1.000   &  &  \cellcolor{red!10} 0.058  & \cellcolor{red!10} 0.101  & \cellcolor{red!10} 0.339  & \cellcolor{red!10} 0.895 & \cellcolor{red!10} 1.000    \\
    & {\tt NET($\lambda=1.6$)}  & \cellcolor{red!7} 0.054  & \cellcolor{red!7} 0.111  & \cellcolor{red!7} 0.372  & \cellcolor{red!7} 0.881 & \cellcolor{red!7} 1.000   &  &  \cellcolor{red!13} 0.038  & \cellcolor{red!13} 0.087  & \cellcolor{red!13} 0.389  & \cellcolor{red!13} 0.877 & \cellcolor{red!13} 1.000     \\
   &{\tt SRM-A}  & \cellcolor{black!20} $\times$  &  \cellcolor{black!20} $\times$  &  \cellcolor{black!20} $\times$  &  \cellcolor{black!20} $\times$ & \cellcolor{black!20} $\times$   &  &   \cellcolor{black!20} $\times$  &  \cellcolor{black!20} $\times$  &  \cellcolor{black!20} $\times$  &  \cellcolor{black!20} $\times$ & \cellcolor{black!20} $\times$    \\
    &{\tt SRM-L} &  \cellcolor{black!20} $\times$  &  \cellcolor{black!20} $\times$  &  \cellcolor{black!20} $\times$  &  \cellcolor{black!20} $\times$ & \cellcolor{black!20} $\times$   &  &   \cellcolor{black!20} $\times$  &  \cellcolor{black!20} $\times$  &  \cellcolor{black!20} $\times$  &  \cellcolor{black!20} $\times$ & \cellcolor{black!20} $\times$     \\
    \midrule
    $n=100$ & {\tt NET($\lambda=1$)}  & \cellcolor{blue!2} 0.052  & \cellcolor{blue!2} 0.155  & \cellcolor{blue!2} 0.659  & \cellcolor{blue!2} 0.995 & \cellcolor{blue!2} 1.000   &  &  \cellcolor{red!5} 0.053  & \cellcolor{red!5} 0.165  & \cellcolor{red!5} 0.655  & \cellcolor{red!5} 0.998 & \cellcolor{red!5} 1.000    \\
    & {\tt NET($\lambda=1.2$)}  & \cellcolor{blue!5} 0.041  & \cellcolor{blue!5} 0.156  & \cellcolor{blue!5} 0.674  & \cellcolor{blue!5} 0.993 & \cellcolor{blue!5} 1.000   &  &  \cellcolor{blue!2} 0.055  & \cellcolor{blue!2} 0.125  & \cellcolor{blue!2} 0.658  & \cellcolor{blue!2} 0.997 & \cellcolor{blue!2} 1.000     \\
    & {\tt NET($\lambda=1.6$)}  & \cellcolor{red!1} 0.074  & \cellcolor{red!1} 0.164  & \cellcolor{red!1} 0.667  & \cellcolor{red!1} 0.996 & \cellcolor{red!1} 1.000   &  &  \cellcolor{red!15} 0.052  & \cellcolor{red!15} 0.152  & \cellcolor{red!15} 0.651  & \cellcolor{red!15} 0.999 & \cellcolor{red!15} 1.000    \\
    &{\tt SRM-A}  & \cellcolor{black!20} $\times$  &  \cellcolor{black!20} $\times$  &  \cellcolor{black!20} $\times$  &  \cellcolor{black!20} $\times$ & \cellcolor{black!20} $\times$   &  &   \cellcolor{black!20} $\times$  &  \cellcolor{black!20} $\times$  &  \cellcolor{black!20} $\times$  &  \cellcolor{black!20} $\times$ & \cellcolor{black!20} $\times$    \\
    &{\tt SRM-L} &  \cellcolor{black!20} $\times$  &  \cellcolor{black!20} $\times$  &  \cellcolor{black!20} $\times$  &  \cellcolor{black!20} $\times$ & \cellcolor{black!20} $\times$   &  &   \cellcolor{black!20} $\times$  &  \cellcolor{black!20} $\times$  &  \cellcolor{black!20} $\times$  &  \cellcolor{black!20} $\times$ & \cellcolor{black!20} $\times$    \\ \midrule
    && \multicolumn{5}{c}{$\eqref{test_comp_a}$ - Normal configuration}          &  & \multicolumn{5}{c}{$\eqref{test_comp_a}$ - Poisson configuration}      \\ \cmidrule{3-7} \cmidrule{9-13} 
    & $c^2$ & 0.05 & 0.2 & 0.5 & 1 & 5 &  & 0.05 & 0.2 & 0.5 & 1 & 5 \\ \midrule
    $n=50$ & {\tt NET($\lambda=1$)}  & \cellcolor{red!4} 0.144  & \cellcolor{red!4} 0.743  & \cellcolor{red!4} 0.999  & \cellcolor{red!4} 1.000 & \cellcolor{red!4} 1.000   &  &  \cellcolor{blue!3} 0.142  & \cellcolor{blue!3} 0.736 & \cellcolor{blue!3} 0.995  & \cellcolor{blue!3} 1.000 & \cellcolor{blue!3} 1.000    \\
    & {\tt NET($\lambda=1.2$)}  & \cellcolor{red!2} 0.232  & \cellcolor{red!2} 0.958  & \cellcolor{red!2} 1.000  & \cellcolor{red!2} 1.000 & \cellcolor{red!2} 1.000   &  &  \cellcolor{red!11} 0.237  & \cellcolor{red!11} 0.946  & \cellcolor{red!11} 1.000  & \cellcolor{red!11} 1.000 & \cellcolor{red!11} 1.000    \\
    & {\tt NET($\lambda=1.6$)}  & \cellcolor{red!15} 0.727  & \cellcolor{red!15} 1.000  & \cellcolor{red!15} 1.000  & \cellcolor{red!15} 1.000 & \cellcolor{red!15} 1.000   &  &  \cellcolor{blue!3} 0.734  & \cellcolor{blue!3} 1.000  & \cellcolor{blue!3} 1.000  & \cellcolor{blue!3} 1.000 & \cellcolor{blue!3} 1.000     \\
    &{\tt SRM-A}  & \cellcolor{blue!23} 0.998  & \cellcolor{blue!23} 1.000  & \cellcolor{blue!23} 1.000  & \cellcolor{blue!23} 1.000 & \cellcolor{blue!23} 1.000   &  & \cellcolor{blue!24} 1.000  & \cellcolor{blue!24} 1.000  & \cellcolor{blue!24} 1.000  & \cellcolor{blue!24} 1.000 & \cellcolor{blue!24} 1.000    \\
    &{\tt SRM-L} & \cellcolor{red!50.1} 1.000  & \cellcolor{red!50.1} 1.000  & \cellcolor{red!50.1} 1.000  & \cellcolor{red!50.1} 1.000 & \cellcolor{red!50.1} 1.000   &  & \cellcolor{red!50.5} 0.997  & \cellcolor{red!50.5} 1.000  & \cellcolor{red!50.5} 1.000  & \cellcolor{red!50.5} 1.000 & \cellcolor{red!50.5} 1.000    \\
    \midrule
    $n=100$ & {\tt NET($\lambda=1$)}  & \cellcolor{blue!0} 0.224  & \cellcolor{blue!0} 0.953  & \cellcolor{blue!0} 1.000  & \cellcolor{blue!0} 1.000 & \cellcolor{blue!0} 1.000   &  &  \cellcolor{blue!11} 0.234  & \cellcolor{blue!11} 0.951  & \cellcolor{blue!11} 1.000  & \cellcolor{blue!11} 1.000 & \cellcolor{blue!11} 1.000    \\
    & {\tt NET($\lambda=1.2$)}  & \cellcolor{red!10} 0.444  & \cellcolor{red!10} 0.999  & \cellcolor{red!10} 1.000  & \cellcolor{red!10} 1.000 & \cellcolor{red!10} 1.000   &  &  \cellcolor{blue!5} 0.466  & \cellcolor{blue!5} 0.999  & \cellcolor{blue!5} 1.000  & \cellcolor{blue!5} 1.000 & \cellcolor{blue!5} 1.000     \\
    & {\tt NET($\lambda=1.6$)}  & \cellcolor{blue!3} 0.977  & \cellcolor{blue!3} 1.000  & \cellcolor{blue!3} 1.000  & \cellcolor{blue!3} 1.000 & \cellcolor{blue!3} 1.000   &  &  \cellcolor{blue!0} 0.989  & \cellcolor{blue!0} 1.000  & \cellcolor{blue!0} 1.000  & \cellcolor{blue!0} 1.000 & \cellcolor{blue!0} 1.000    \\
    &{\tt SRM-A}  & \cellcolor{blue!3} 1.000  & \cellcolor{blue!3} 1.000  & \cellcolor{blue!3} 1.000  & \cellcolor{blue!3} 1.000 &  \cellcolor{blue!3} 1.000   &  &  \cellcolor{blue!4} 1.000  & \cellcolor{blue!4} 1.000  & \cellcolor{blue!4} 1.000  & \cellcolor{blue!4} 1.000 & \cellcolor{blue!4} 1.000    \\
    &{\tt SRM-L} &  \cellcolor{black!20} $\times$  &  \cellcolor{black!20} $\times$  &  \cellcolor{black!20} $\times$  &  \cellcolor{black!20} $\times$ & \cellcolor{black!20} $\times$   &  &   \cellcolor{black!20} $\times$  &  \cellcolor{black!20} $\times$  &  \cellcolor{black!20} $\times$  &  \cellcolor{black!20} $\times$ & \cellcolor{black!20} $\times$    \\ \midrule
    && \multicolumn{5}{c}{$\eqref{test_comp_d}$ - Normal configuration}          &  & \multicolumn{5}{c}{$\eqref{test_comp_d}$ - Poisson configuration}      \\ \cmidrule{3-7} \cmidrule{9-13} 
    & $c^2$ & 0.05 & 0.2 & 0.5 & 1 & 5 &  & 0.05 & 0.2 & 0.5 & 1 & 5 \\ \midrule
    $n=50$ & {\tt NET($\lambda=1$)}  & \cellcolor{blue!9} 0.170  & \cellcolor{blue!9} 0.624  & \cellcolor{blue!9} 0.980  & \cellcolor{blue!9} 0.995 & \cellcolor{blue!9} 0.995    &  & \cellcolor{red!15} 0.155  &\cellcolor{red!15} 0.626  & \cellcolor{red!15} 0.957  & \cellcolor{red!15} 0.969 & \cellcolor{red!15} 0.972    \\
    & {\tt NET($\lambda=1.2$)}  & \cellcolor{blue!4} 0.267  & \cellcolor{blue!4} 0.833  & \cellcolor{blue!4} 0.993  & \cellcolor{blue!4} 0.998 & \cellcolor{blue!4} 0.998    &  &  \cellcolor{red!12} 0.272  & \cellcolor{red!12} 0.827  & \cellcolor{red!12} 0.969  & \cellcolor{red!12} 0.972 & \cellcolor{red!12} 0.975    \\
    & {\tt NET($\lambda=1.6$)}  & \cellcolor{red!33} 0.722  & \cellcolor{red!33} 0.991  & \cellcolor{red!33} 0.997  & \cellcolor{red!33} 0.994 & \cellcolor{red!33} 0.998  &  &  \cellcolor{red!27} 0.683  & \cellcolor{red!27} 0.982  & \cellcolor{red!27} 0.966  & \cellcolor{red!27} 0.989 & \cellcolor{red!27} 0.977    \\
    &{\tt SRM-A}  & \cellcolor{red!73.8}  0.998  & \cellcolor{red!73.8} 1.000  & \cellcolor{red!73.8} 1.000  & \cellcolor{red!73.8} 1.000 & \cellcolor{red!73.8} 1.000   &  & \cellcolor{red!75.7}  0.993  & \cellcolor{red!75.7} 1.000  & \cellcolor{red!75.7} 1.000  & \cellcolor{red!75.7} 1.000 & \cellcolor{red!75.7} 1.000    \\
    &{\tt SRM-L} &  \cellcolor{red!83.7}  0.999  & \cellcolor{red!83.7} 1.000  & \cellcolor{red!83.7} 1.000  & \cellcolor{red!83.7} 1.000 & \cellcolor{red!83.7} 1.000  &  & \cellcolor{red!83.9}  0.993  & \cellcolor{red!83.9} 1.000  & \cellcolor{red!83.9} 1.000  & \cellcolor{red!83.9} 1.000 & \cellcolor{red!83.9} 1.000    \\
    \midrule
    $n=100$ & {\tt NET($\lambda=1$)}  & \cellcolor{blue!3} 0.265  & \cellcolor{blue!3} 0.860  &  \cellcolor{blue!3} 1.000  & \cellcolor{blue!3} 1.000 & \cellcolor{blue!3} 1.000   &  &  \cellcolor{blue!12} 0.244  & \cellcolor{blue!12} 0.897  & \cellcolor{blue!12} 0.998  & \cellcolor{blue!12} 1.000 & \cellcolor{blue!12} 1.000    \\
    & {\tt NET($\lambda=1.2$)}  & \cellcolor{blue!3} 0.466  & \cellcolor{blue!3} 0.989  & \cellcolor{blue!3} 1.000  & \cellcolor{blue!3} 1.000 & \cellcolor{blue!3} 1.000  &  &  \cellcolor{blue!10} 0.456  & \cellcolor{blue!10} 0.982  & \cellcolor{blue!10} 1.000  & \cellcolor{blue!10} 1.000 & \cellcolor{blue!10} 1.000    \\
    & {\tt NET($\lambda=1.6$)}  &  \cellcolor{red!33} 0.976  & \cellcolor{red!33} 1.000  & \cellcolor{red!33} 1.000  & \cellcolor{red!33} 1.000 & \cellcolor{red!33} 1.000  &  &  \cellcolor{red!11} 0.947  & \cellcolor{red!11} 1.000  & \cellcolor{red!11} 1.000  & \cellcolor{red!11} 1.000 & \cellcolor{red!11} 1.000    \\
    &{\tt SRM-A}  &  \cellcolor{red!85.3} 1.000  & \cellcolor{red!85.3} 1.000  & \cellcolor{red!85.3} 1.000  & \cellcolor{red!85.3} 1.000 & \cellcolor{red!85.3} 1.000  &  &  \cellcolor{red!86.9} 1.000  & \cellcolor{red!86.9} 1.000  & \cellcolor{red!86.9} 1.000  & \cellcolor{red!86.9} 1.000 & \cellcolor{red!86.9} 1.000    \\
    &{\tt SRM-L} &  \cellcolor{black!20} $\times$  &  \cellcolor{black!20} $\times$  &  \cellcolor{black!20} $\times$  &  \cellcolor{black!20} $\times$ & \cellcolor{black!20} $\times$   &  &   \cellcolor{black!20} $\times$  &  \cellcolor{black!20} $\times$  &  \cellcolor{black!20} $\times$  &  \cellcolor{black!20} $\times$ & \cellcolor{black!20} $\times$    \\
    \bottomrule
     \multicolumn{13}{l}{
     \begin{minipage}{0.28\textwidth}
        \vspace{-5pt}
            Background color: type-I error rate (see Table \ref{tab::smry_size_marginal_C=1_updated}), color key:
        \end{minipage}
     \begin{minipage}{0.6\textwidth}
     \vspace{5pt}
        \begin{tikzpicture}
      \pgfdeclarehorizontalshading{myshading}{200bp}
      {
        color(0bp)=(blue!100);
        color(7.5bp)=(blue!25);
        color(15bp)=(white);
        color(15bp)=(white);
        color(22.5bp)=(red!25);
        color(30bp)=(red!50);
        color(50bp)=(red!100)
      }
      \path[shading=myshading] (0,0) rectangle (10.2,0.25);
          \node at (0, -0.35) {0};
          \node at (3.06, -0.35) {\bf 0.05};
          \node at (10.2, -0.35) {1};
          \draw (0.01, -0.1) -- (0.01, 0);
          \draw (3.06, -0.1) -- (3.06, 0.1);
          \draw (10.19, -0.1) -- (10.19, 0.1);
    \end{tikzpicture}
     \end{minipage}
     }   
    \end{tabular}
    \end{center}
\end{table}

As anticipated, the empirical power of {\tt NET} rapidly approaches $1$ with increasing network effect signal $c$, particularly evident when $c\geq 1$ across all settings, thereby affirming our method's consistency. When employing reduced network moments in tests, our method's power increases with a higher $\lambda$ value. Notably, $\lambda=1.2$ strikes a balance between satisfactory power and fair type-I error rate control. In setting \eqref{test_comp_a}, where {\tt SRM-A} experiences no model misspecifications, its strong performance in both power and type-I error rate control is expected. However, under setting \eqref{test_comp_d}, where the true model deviates from the additive assumption, {\tt SRM-A}'s elevated power becomes inconsequential and misleading due to significant inflation in the type-I error rate. Similar patterns are observed for {\tt SRM-L}.

\section{Data example on faculty hiring networks}
\label{section::real-data}

In this section, we employ our proposed method to analyze the faculty hiring networks \citep{clauset2015facultydata}. For the diagnosis of degeneracy when testing the reciprocity and sender-receiver effects, we set the prespecified $C=1$, as detailed in Sections \ref{subsubsec::moment-based-test-of-network-effects::test-sender-receiver-effect::test-degeneracy} and \ref{subsubsec::moment-based-test-of-network-effects::test-reciprocity-effect::test-degeneracy}. Guided by our simulation results, we chose $\lambda = 1.2$ to balance between maintaining an accurate type-I error rate control and achieving satisfactory power. An additional data example on the  bilateral trade flows with trade-related variables is documented in the Supplementary Section F.2.
    
The faculty hiring networks, collected by \cite{clauset2015facultydata}, aimed to investigate the institutional prestige ranking that best explains the observed faculty hiring patterns in computer science, business schools, and history. In these networks for the three distinct disciplines, each node represents a Ph.D.-granting institution in its respective field. A directed edge from node $i$ to node $j$ signifies that an individual earned a Ph.D. from institution $i$ and held a tenure-track faculty position at institution $j$ at the time of data collection (2010 for computer science, 2012 for business, and 2009 for history). Our analysis focuses on the number of faculty positions filled by institution $i$ from institution $j$, denoted by the count outcome $e_{i,j}$, while excluding self-loops. For data under considerations, the computer science hiring network includes $205$ and $2,881$ directed edges with positive count weights. The business school hiring network features $112$ nodes and $3,404$ directed edges with positive count weights. Lastly, the history hiring network encompasses $144$ nodes and $2,372$ directed edges, each carrying positive count weights.
    
For visualization, we consider a local version of network effects for each node. For node $i \in [n]$, the local reciprocity effect is calculated by $(n-1)^{-1}\sum_{j \in [n], j\neq i} (e_{i,j} - \hat\mu)(e_{j,i} - \hat\mu)$; the local same-sender effect is computed through $(n-1)^{-1} (n-2)^{-1}\sum_{j,k: (i,j,k)\textrm{ distinct}} (e_{i,j} - \hat\mu)(e_{i,k} - \hat\mu)$; and the local same-receiver effect and local sender-receiver effect are similarly derived by replacing $(e_{i,j} - \hat\mu)(e_{i,k} - \hat\mu)$ with $(e_{j,i} - \hat\mu)(e_{k,i} - \hat\mu)$ and $(e_{j,i} - \hat\mu)(e_{i,k} - \hat\mu)$, respectively, with shorthand $\hat\mu$ denoting the sample mean of $\{e_{i,j}\}_{1 \leq \{i,j\} \leq n}$.
    
\begin{figure}[h!]
\centering
    \subfigure{
    \includegraphics[width=0.51\textwidth]{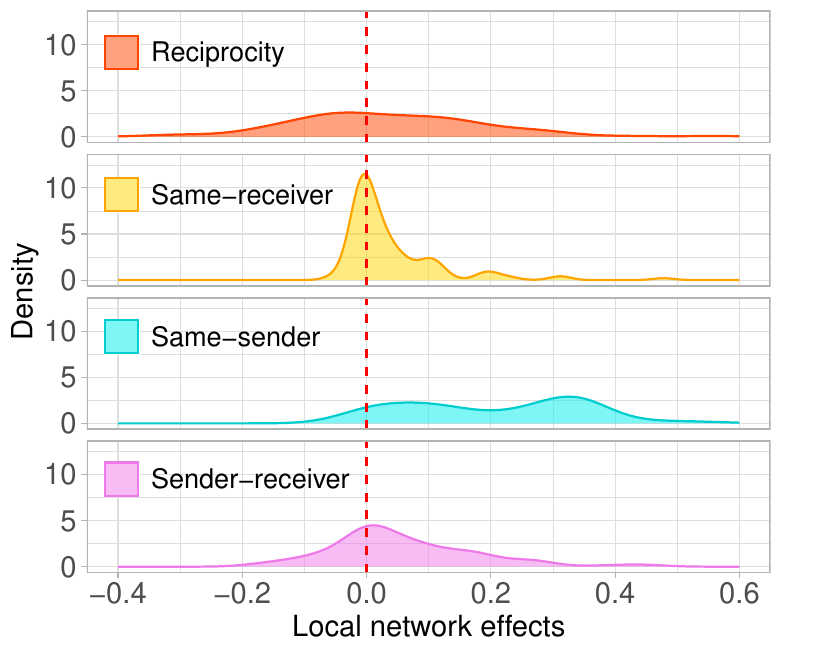}}
    \subfigure{
    \includegraphics[width=0.43\textwidth]{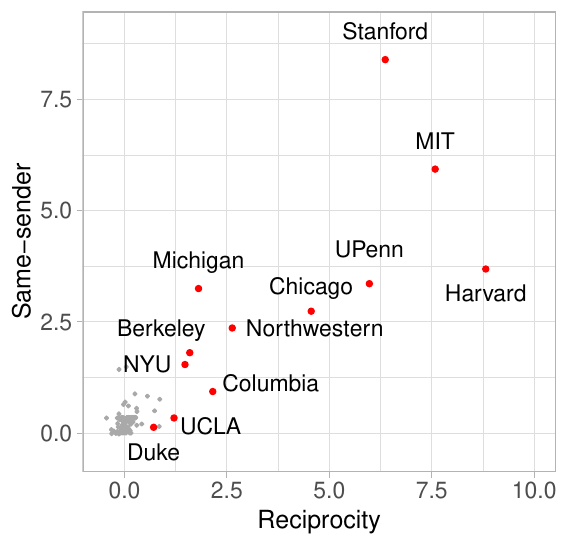}}
    \caption{Local network effects for the business school hiring network. Left: Density plots of local network effects. Right: Scatter plot of the local reciprocity effect versus the local same-sender effect. Institutes with larger local reciprocity and same-sender effects are highlighted in red.}
\label{fig::local-network-effects}
\end{figure}

\begin{figure}[h!]
    \centering
    \includegraphics[width=\linewidth]{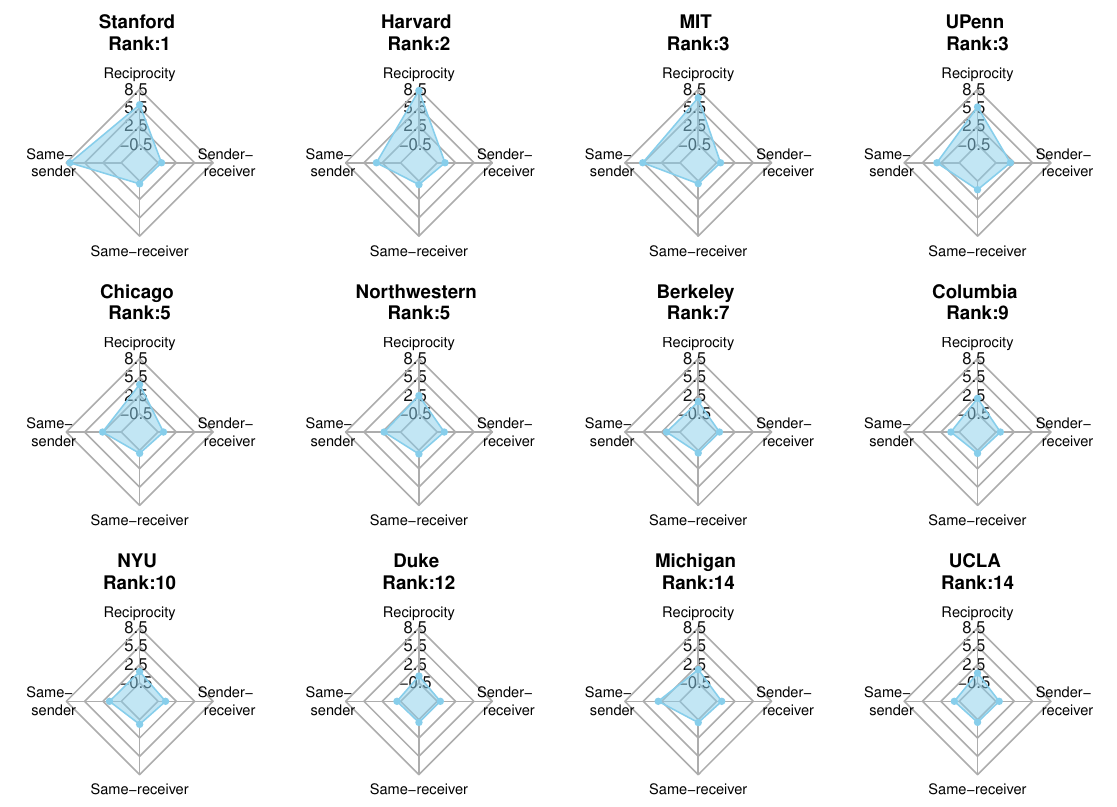} 
    \caption {Radar plots of local network effects for institutions marked in red in Figure \ref{fig::local-network-effects}, ordered by the \textit{U.S. News \& World Report} rankings in 2012 for business schools.} \label{fig::radar-indi-school-1} 
\end{figure}   

Figure \ref{fig::local-network-effects} presents the density and scatter plots of local network effects in the business school hiring network. The density plot reveals a striking disparity in the distributions of the local same-receiver and same-sender effects. The local same-receiver effects, likely influenced by hiring mechanisms, show less variation across institutions, implying that institutions adopt similar hiring strategies. In contrast, the same-sender effect, indicative of an institution's success in producing graduates who are in demand by other institutions, appears to vary substantially among institutions. The scatter plot highlights the local reciprocity effect against the same-sender effect, while the other two effects are relatively minor (close to $0$). Furthermore, Figure \ref{fig::radar-indi-school-1} clearly illustrates that top-ranked institutions show more pronounced reciprocity and same-sender effects compared to their lower-ranked counterparts. This aligns with the observation that institutions of all tiers endeavor to recruit the best candidates, with the majority tending to consider graduates from institutions of higher or similar tiers, while top-tier institutions often engage in reciprocal hiring.
    
\begin{table}[h!] 
\label{tab::FacultyData_MF_alt}
    \caption {The $p$-values for testing distinct network effects across disciplines.} 
    \begin{center}
    \begin{tabular}{lcccc}
    \toprule
    & $H_0:\eta_2=0$ & $H_0:\eta_3=0$ & $H_0:\eta_4=0$ & $H_0:\eta_5=0$ \\
    &  (reciprocity) &
    (same-sender) &  (same-receiver) & (sender-receiver)\\
   \midrule
    Business  
    & $0.102$ & $\bm{<10^{-3}}$  &  $0.418$ &  $0.093$\\ History & 
    $\textbf{0.021}$ & $\bm{<10^{-3}}$ & $0.558$ &     $\textbf{0.042}$ \\
     Computer Science &    $\textbf{0.029}$ &  $\bm{<10^{-3}}$ & $0.556$  &$0.095$  \\
    \bottomrule
    \end{tabular}
    \end{center}
\end{table}
     
Table \ref{tab::FacultyData_MF_alt} reports the $p$-values from our tests on various network effects. For tests employing reduced network moments, we generated $10,000$ repeated random sampling. The resulting $p$-values were aggregated using the Z-average test \citep{liu2022multiple}, a combination test that ensures the type-I error rate control under general dependence \citep{diciccio2019multiple, liu2020cauchy, zhang2021beauty}. Additional details are available in the Supplementary Section F.1. Our method detects significant same-sender effects across all three fields, aligning with the patterns observed in the density plots for the business school hiring network in Figure \ref{fig::local-network-effects}. This result highlights the presence of heterogeneity in institutions' capacities to produce competitive candidates \citep{zeleneev2020identification,candelaria2020semiparametric,chen2021trade,johnsson2021estimation}. Furthermore, Table \ref{tab::FacultyData_MF_alt} reveals reciprocal hiring behaviors within the disciplines of computer science and history, coupled with a pronounced sender-receiver effect among history departments. This may suggest that institutions recruiting from higher-tier counterparts potentially enhance their output of competitive candidates, thereby reinforcing their role as influential senders.


\bibliographystyle{imsart-nameyear}
\bibliography{mybibliography}

\begin{thebibliography}{93}

\bibitem[\protect\citeauthoryear{Aldous}{1981}]{aldous1981representations}
\begin{barticle}[author]
\bauthor{\bsnm{Aldous},~\bfnm{David~J}\binits{D.~J.}}
(\byear{1981}).
\btitle{Representations for partially exchangeable arrays of random variables}.
\bjournal{Journal of Multivariate Analysis}
\bvolume{11}
\bpages{581--598}.
\end{barticle}
\endbibitem

\bibitem[\protect\citeauthoryear{Aldous}{1985}]{aldous1985exchangeability}
\begin{bincollection}[author]
\bauthor{\bsnm{Aldous},~\bfnm{David~J}\binits{D.~J.}}
(\byear{1985}).
\btitle{Exchangeability and related topics}.
In \bbooktitle{{\'E}cole d'{\'E}t{\'e} de Probabilit{\'e}s de Saint-Flour XIII—1983}
\bpages{1--198}.
\bpublisher{Springer}.
\end{bincollection}
\endbibitem

\bibitem[\protect\citeauthoryear{Anderson}{2011}]{anderson2011gravity}
\begin{barticle}[author]
\bauthor{\bsnm{Anderson},~\bfnm{James~E}\binits{J.~E.}}
(\byear{2011}).
\btitle{The gravity model}.
\bjournal{Annual Review of Economics}
\bvolume{3}
\bpages{133--160}.
\end{barticle}
\endbibitem

\bibitem[\protect\citeauthoryear{Behar and Nelson}{2014}]{behar2014trade}
\begin{barticle}[author]
\bauthor{\bsnm{Behar},~\bfnm{Alberto}\binits{A.}} \AND \bauthor{\bsnm{Nelson},~\bfnm{Benjamin~D}\binits{B.~D.}}
(\byear{2014}).
\btitle{Trade flows, multilateral resistance, and firm heterogeneity}.
\bjournal{Review of Economics and Statistics}
\bvolume{96}
\bpages{538--549}.
\end{barticle}
\endbibitem

\bibitem[\protect\citeauthoryear{Bickel, Chen and Levina}{2011}]{bickel2011method}
\begin{barticle}[author]
\bauthor{\bsnm{Bickel},~\bfnm{Peter~J}\binits{P.~J.}}, \bauthor{\bsnm{Chen},~\bfnm{Aiyou}\binits{A.}} \AND \bauthor{\bsnm{Levina},~\bfnm{Elizaveta}\binits{E.}}
(\byear{2011}).
\btitle{The method of moments and degree distributions for network models}.
\bjournal{The Annals of Statistics}
\bvolume{39}
\bpages{2280--2301}.
\end{barticle}
\endbibitem

\bibitem[\protect\citeauthoryear{Bond and Lashley}{1996}]{bond1996round}
\begin{barticle}[author]
\bauthor{\bsnm{Bond},~\bfnm{Charles~F}\binits{C.~F.}} \AND \bauthor{\bsnm{Lashley},~\bfnm{Brian~R}\binits{B.~R.}}
(\byear{1996}).
\btitle{Round-robin analysis of social interaction: Exact and estimated standard errors}.
\bjournal{Psychometrika}
\bvolume{61}
\bpages{303--311}.
\end{barticle}
\endbibitem

\bibitem[\protect\citeauthoryear{Borgs, Chayes and Lov{\'a}sz}{2010}]{borgs2010moments}
\begin{barticle}[author]
\bauthor{\bsnm{Borgs},~\bfnm{Christian}\binits{C.}}, \bauthor{\bsnm{Chayes},~\bfnm{Jennifer}\binits{J.}} \AND \bauthor{\bsnm{Lov{\'a}sz},~\bfnm{L{\'a}szl{\'o}}\binits{L.}}
(\byear{2010}).
\btitle{Moments of two-variable functions and the uniqueness of graph limits}.
\bjournal{Geometric and Functional Analysis}
\bvolume{19}
\bpages{1597--1619}.
\end{barticle}
\endbibitem

\bibitem[\protect\citeauthoryear{Cai and Ma}{2013}]{cai2013optimal}
\begin{barticle}[author]
\bauthor{\bsnm{Cai},~\bfnm{T~Tony}\binits{T.~T.}} \AND \bauthor{\bsnm{Ma},~\bfnm{Zongming}\binits{Z.}}
(\byear{2013}).
\btitle{Optimal hypothesis testing for high dimensional covariance matrices}.
\bjournal{Bernoulli}
\bvolume{19}
\bpages{2359--2388}.
\end{barticle}
\endbibitem

\bibitem[\protect\citeauthoryear{Candelaria}{2020}]{candelaria2020semiparametric}
\begin{barticle}[author]
\bauthor{\bsnm{Candelaria},~\bfnm{Luis~E}\binits{L.~E.}}
(\byear{2020}).
\btitle{A semiparametric network formation model with unobserved linear heterogeneity}.
\bjournal{arXiv preprint arXiv:2007.05403}.
\end{barticle}
\endbibitem

\bibitem[\protect\citeauthoryear{Card et~al.}{2005}]{card2005gender}
\begin{barticle}[author]
\bauthor{\bsnm{Card},~\bfnm{Noel}\binits{N.}}, \bauthor{\bsnm{Hodges},~\bfnm{Ernest}\binits{E.}}, \bauthor{\bsnm{Little},~\bfnm{Todd}\binits{T.}} \AND \bauthor{\bsnm{Hawley},~\bfnm{Patricia}\binits{P.}}
(\byear{2005}).
\btitle{Gender effects in peer nominations for aggression and social status}.
\bjournal{International Journal of Behavioral Development}
\bvolume{29}
\bpages{146--155}.
\end{barticle}
\endbibitem

\bibitem[\protect\citeauthoryear{Chandrasekhar}{2016}]{chandrasekhar2016cid}
\begin{barticle}[author]
\bauthor{\bsnm{Chandrasekhar},~\bfnm{Arun}\binits{A.}}
(\byear{2016}).
\btitle{Econometrics of network formation}.
\bjournal{The Oxford Handbook of the Economics of Networks}
\bpages{303--357}.
\end{barticle}
\endbibitem

\bibitem[\protect\citeauthoryear{Chen, Fern{\'a}ndez-Val and Weidner}{2021}]{chen2021trade}
\begin{barticle}[author]
\bauthor{\bsnm{Chen},~\bfnm{Mingli}\binits{M.}}, \bauthor{\bsnm{Fern{\'a}ndez-Val},~\bfnm{Iv{\'a}n}\binits{I.}} \AND \bauthor{\bsnm{Weidner},~\bfnm{Martin}\binits{M.}}
(\byear{2021}).
\btitle{Nonlinear factor models for network and panel data}.
\bjournal{Journal of Econometrics}
\bvolume{220}
\bpages{296--324}.
\end{barticle}
\endbibitem

\bibitem[\protect\citeauthoryear{Chen and Kato}{2019}]{Chen2019U}
\begin{barticle}[author]
\bauthor{\bsnm{Chen},~\bfnm{Xiaohui}\binits{X.}} \AND \bauthor{\bsnm{Kato},~\bfnm{Kengo}\binits{K.}}
(\byear{2019}).
\btitle{Randomized incomplete U-statistics in high dimensions}.
\bjournal{The Annals of Statistics}
\bvolume{47}
\bpages{3127--3156}.
\end{barticle}
\endbibitem

\bibitem[\protect\citeauthoryear{Clauset, Arbesman and Larremore}{2015}]{clauset2015facultydata}
\begin{barticle}[author]
\bauthor{\bsnm{Clauset},~\bfnm{Aaron}\binits{A.}}, \bauthor{\bsnm{Arbesman},~\bfnm{Samuel}\binits{S.}} \AND \bauthor{\bsnm{Larremore},~\bfnm{Daniel~B}\binits{D.~B.}}
(\byear{2015}).
\btitle{Systematic inequality and hierarchy in faculty hiring networks}.
\bjournal{Science Advances}
\bvolume{1}
\bpages{e1400005}.
\end{barticle}
\endbibitem

\bibitem[\protect\citeauthoryear{Cockerham and Weir}{1977}]{cockerham1977quadratic}
\begin{barticle}[author]
\bauthor{\bsnm{Cockerham},~\bfnm{C~Clark}\binits{C.~C.}} \AND \bauthor{\bsnm{Weir},~\bfnm{Bruce~Spencer}\binits{B.~S.}}
(\byear{1977}).
\btitle{Quadratic analyses of reciprocal crosses}.
\bjournal{Biometrics}
\bpages{187--203}.
\end{barticle}
\endbibitem

\bibitem[\protect\citeauthoryear{Cranmer, Heinrich and Desmarais}{2014}]{cranmer2014reciprocityEffect}
\begin{barticle}[author]
\bauthor{\bsnm{Cranmer},~\bfnm{Skyler~J}\binits{S.~J.}}, \bauthor{\bsnm{Heinrich},~\bfnm{Tobias}\binits{T.}} \AND \bauthor{\bsnm{Desmarais},~\bfnm{Bruce~A}\binits{B.~A.}}
(\byear{2014}).
\btitle{Reciprocity and the structural determinants of the international sanctions network}.
\bjournal{Social Networks}
\bvolume{36}
\bpages{5--22}.
\end{barticle}
\endbibitem

\bibitem[\protect\citeauthoryear{DiCiccio and Romano}{2019}]{diciccio2019multiple}
\begin{btechreport}[author]
\bauthor{\bsnm{DiCiccio},~\bfnm{C}\binits{C.}} \AND \bauthor{\bsnm{Romano},~\bfnm{J}\binits{J.}}
(\byear{2019}).
\btitle{Multiple data splitting for testing}
\btype{Technical Report},
\bpublisher{Technical Report, Department of Statistics, Stanford University}.
\end{btechreport}
\endbibitem

\bibitem[\protect\citeauthoryear{Drton, Han and Shi}{2020}]{drton2020high}
\begin{barticle}[author]
\bauthor{\bsnm{Drton},~\bfnm{Mathias}\binits{M.}}, \bauthor{\bsnm{Han},~\bfnm{Fang}\binits{F.}} \AND \bauthor{\bsnm{Shi},~\bfnm{Hongjian}\binits{H.}}
(\byear{2020}).
\btitle{High-dimensional consistent independence testing with maxima of rank correlations}.
\bjournal{The Annals of Statistics}
\bvolume{48}
\bpages{3206--3227}.
\end{barticle}
\endbibitem

\bibitem[\protect\citeauthoryear{Eisenkraft and Elfenbein}{2010}]{eisenkraft2010way}
\begin{barticle}[author]
\bauthor{\bsnm{Eisenkraft},~\bfnm{Noah}\binits{N.}} \AND \bauthor{\bsnm{Elfenbein},~\bfnm{Hillary~Anger}\binits{H.~A.}}
(\byear{2010}).
\btitle{The way you make me feel: Evidence for individual differences in affective presence}.
\bjournal{Psychological Science}
\bvolume{21}
\bpages{505--510}.
\end{barticle}
\endbibitem

\bibitem[\protect\citeauthoryear{Fafchamps and Gubert}{2007}]{fafchamps2007formation}
\begin{barticle}[author]
\bauthor{\bsnm{Fafchamps},~\bfnm{Marcel}\binits{M.}} \AND \bauthor{\bsnm{Gubert},~\bfnm{Flore}\binits{F.}}
(\byear{2007}).
\btitle{The formation of risk sharing networks}.
\bjournal{Journal of Development Economics}
\bvolume{83}
\bpages{326--350}.
\end{barticle}
\endbibitem

\bibitem[\protect\citeauthoryear{Fan and Li}{1999}]{fan1999central}
\begin{barticle}[author]
\bauthor{\bsnm{Fan},~\bfnm{Yanqin}\binits{Y.}} \AND \bauthor{\bsnm{Li},~\bfnm{Qi}\binits{Q.}}
(\byear{1999}).
\btitle{Central limit theorem for degenerate U-statistics of absolutely regular processes with applications to model specification testing}.
\bjournal{Journal of Nonparametric Statistics}
\bvolume{10}
\bpages{245--271}.
\end{barticle}
\endbibitem

\bibitem[\protect\citeauthoryear{Fan et~al.}{2022a}]{fan2022simple}
\begin{barticle}[author]
\bauthor{\bsnm{Fan},~\bfnm{Jianqing}\binits{J.}}, \bauthor{\bsnm{Fan},~\bfnm{Yingying}\binits{Y.}}, \bauthor{\bsnm{Han},~\bfnm{Xiao}\binits{X.}} \AND \bauthor{\bsnm{Lv},~\bfnm{Jinchi}\binits{J.}}
(\byear{2022}a).
\btitle{SIMPLE: Statistical inference on membership profiles in large networks}.
\bjournal{Journal of the Royal Statistical Society Series B: Statistical Methodology}
\bvolume{84}
\bpages{630--653}.
\end{barticle}
\endbibitem

\bibitem[\protect\citeauthoryear{Fan et~al.}{2022b}]{fan2023simpleRC}
\begin{barticle}[author]
\bauthor{\bsnm{Fan},~\bfnm{Jianqing}\binits{J.}}, \bauthor{\bsnm{Fan},~\bfnm{Yingying}\binits{Y.}}, \bauthor{\bsnm{Lv},~\bfnm{Jinchi}\binits{J.}} \AND \bauthor{\bsnm{Yang},~\bfnm{Fan}\binits{F.}}
(\byear{2022}b).
\btitle{SIMPLE-RC: Group network inference with non-sharp nulls and weak signals}.
\bjournal{arXiv preprint arXiv:2211.00128}.
\end{barticle}
\endbibitem

\bibitem[\protect\citeauthoryear{Fisher and Lee}{1983}]{fisher1983correlation}
\begin{barticle}[author]
\bauthor{\bsnm{Fisher},~\bfnm{Nick~I}\binits{N.~I.}} \AND \bauthor{\bsnm{Lee},~\bfnm{Alan~J}\binits{A.~J.}}
(\byear{1983}).
\btitle{A correlation coefficient for circular data}.
\bjournal{Biometrika}
\bvolume{70}
\bpages{327--332}.
\end{barticle}
\endbibitem

\bibitem[\protect\citeauthoryear{Frank and Strauss}{1986}]{frank1986markov}
\begin{barticle}[author]
\bauthor{\bsnm{Frank},~\bfnm{Ove}\binits{O.}} \AND \bauthor{\bsnm{Strauss},~\bfnm{David}\binits{D.}}
(\byear{1986}).
\btitle{Markov graphs}.
\bjournal{Journal of the American Statistical Association}
\bvolume{81}
\bpages{832--842}.
\end{barticle}
\endbibitem

\bibitem[\protect\citeauthoryear{Fredrickson and Chen}{2019}]{fredrickson2019permutation}
\begin{barticle}[author]
\bauthor{\bsnm{Fredrickson},~\bfnm{Mark~M}\binits{M.~M.}} \AND \bauthor{\bsnm{Chen},~\bfnm{Yuguo}\binits{Y.}}
(\byear{2019}).
\btitle{Permutation and randomization tests for network analysis}.
\bjournal{Social Networks}
\bvolume{59}
\bpages{171--183}.
\end{barticle}
\endbibitem

\bibitem[\protect\citeauthoryear{Gao, Lu and Zhou}{2015}]{chao2015rate}
\begin{barticle}[author]
\bauthor{\bsnm{Gao},~\bfnm{Chao}\binits{C.}}, \bauthor{\bsnm{Lu},~\bfnm{Yu}\binits{Y.}} \AND \bauthor{\bsnm{Zhou},~\bfnm{Harrison~H.}\binits{H.~H.}}
(\byear{2015}).
\btitle{Rate-optimal graphon estimation}.
\bjournal{The Annals of Statistics}
\bvolume{43}
\bpages{2624--2652}.
\end{barticle}
\endbibitem

\bibitem[\protect\citeauthoryear{Gao et~al.}{2016}]{gao2016optimal}
\begin{barticle}[author]
\bauthor{\bsnm{Gao},~\bfnm{Chao}\binits{C.}}, \bauthor{\bsnm{Lu},~\bfnm{Yu}\binits{Y.}}, \bauthor{\bsnm{Ma},~\bfnm{Zongming}\binits{Z.}} \AND \bauthor{\bsnm{Zhou},~\bfnm{Harrison~H}\binits{H.~H.}}
(\byear{2016}).
\btitle{Optimal estimation and completion of matrices with biclustering structures}.
\bjournal{The Journal of Machine Learning Research}
\bvolume{17}
\bpages{5602--5630}.
\end{barticle}
\endbibitem

\bibitem[\protect\citeauthoryear{Gelman and Hill}{2006}]{gelman2006ref}
\begin{bbook}[author]
\bauthor{\bsnm{Gelman},~\bfnm{Andrew}\binits{A.}} \AND \bauthor{\bsnm{Hill},~\bfnm{Jennifer}\binits{J.}}
(\byear{2006}).
\btitle{Data analysis using regression and multilevel/hierarchical models}.
\bpublisher{Cambridge University Press}.
\end{bbook}
\endbibitem

\bibitem[\protect\citeauthoryear{Gill and Swartz}{2001}]{gill2001statistical}
\begin{barticle}[author]
\bauthor{\bsnm{Gill},~\bfnm{Paramjit~S}\binits{P.~S.}} \AND \bauthor{\bsnm{Swartz},~\bfnm{Tim~B}\binits{T.~B.}}
(\byear{2001}).
\btitle{Statistical analyses for round robin interaction data}.
\bjournal{Canadian Journal of Statistics}
\bvolume{29}
\bpages{321--331}.
\end{barticle}
\endbibitem

\bibitem[\protect\citeauthoryear{Gin et~al.}{2020}]{gin2020dyadic}
\begin{barticle}[author]
\bauthor{\bsnm{Gin},~\bfnm{Brian}\binits{B.}}, \bauthor{\bsnm{Sim},~\bfnm{Nicholas}\binits{N.}}, \bauthor{\bsnm{Skrondal},~\bfnm{Anders}\binits{A.}} \AND \bauthor{\bsnm{Rabe-Hesketh},~\bfnm{Sophia}\binits{S.}}
(\byear{2020}).
\btitle{A dyadic IRT model}.
\bjournal{Psychometrika}
\bvolume{85}
\bpages{815--836}.
\end{barticle}
\endbibitem

\bibitem[\protect\citeauthoryear{Graham}{2020a}]{graham2020dyadic}
\begin{barticle}[author]
\bauthor{\bsnm{Graham},~\bfnm{Bryan~S}\binits{B.~S.}}
(\byear{2020}a).
\btitle{Dyadic regression}.
\bjournal{The Econometric Analysis of Network Data}
\bpages{23--40}.
\end{barticle}
\endbibitem

\bibitem[\protect\citeauthoryear{Graham}{2020b}]{graham2020cid}
\begin{bincollection}[author]
\bauthor{\bsnm{Graham},~\bfnm{Bryan~S}\binits{B.~S.}}
(\byear{2020}b).
\btitle{Network data}.
In \bbooktitle{Handbook of Econometrics},
\bvolume{7}
\bpages{111--218}.
\bpublisher{Elsevier}.
\end{bincollection}
\endbibitem

\bibitem[\protect\citeauthoryear{Gregory}{1977}]{gregory1977large}
\begin{barticle}[author]
\bauthor{\bsnm{Gregory},~\bfnm{Gavin~G}\binits{G.~G.}}
(\byear{1977}).
\btitle{Large sample theory for U-statistics and tests of fit}.
\bjournal{The Annals of Statistics}
\bpages{110--123}.
\end{barticle}
\endbibitem

\bibitem[\protect\citeauthoryear{Han and Qian}{2018}]{han2018inference}
\begin{barticle}[author]
\bauthor{\bsnm{Han},~\bfnm{Fang}\binits{F.}} \AND \bauthor{\bsnm{Qian},~\bfnm{Tianchen}\binits{T.}}
(\byear{2018}).
\btitle{On inference validity of weighted U-statistics under data heterogeneity}.
\bjournal{Electronic Journal of Statistics}
\bvolume{12}
\bpages{2637--2708}.
\end{barticle}
\endbibitem

\bibitem[\protect\citeauthoryear{Helpman, Melitz and Rubinstein}{2008}]{helpman2008trade}
\begin{barticle}[author]
\bauthor{\bsnm{Helpman},~\bfnm{Elhanan}\binits{E.}}, \bauthor{\bsnm{Melitz},~\bfnm{Marc}\binits{M.}} \AND \bauthor{\bsnm{Rubinstein},~\bfnm{Yona}\binits{Y.}}
(\byear{2008}).
\btitle{Estimating trade flows: Trading partners and trading volumes}.
\bjournal{The Quarterly Journal of Economics}
\bvolume{123}
\bpages{441--487}.
\end{barticle}
\endbibitem

\bibitem[\protect\citeauthoryear{Ho and Shieh}{2006}]{ho2006two}
\begin{barticle}[author]
\bauthor{\bsnm{Ho},~\bfnm{Hwai-chung}\binits{H.-c.}} \AND \bauthor{\bsnm{Shieh},~\bfnm{Grace~S}\binits{G.~S.}}
(\byear{2006}).
\btitle{Two-stage U-statistics for Hypothesis Testing}.
\bjournal{Scandinavian Journal of Statistics}
\bvolume{33}
\bpages{861--873}.
\end{barticle}
\endbibitem

\bibitem[\protect\citeauthoryear{Hoff}{2005}]{Hoff2005mixedEffects}
\begin{barticle}[author]
\bauthor{\bsnm{Hoff},~\bfnm{Peter~D}\binits{P.~D.}}
(\byear{2005}).
\btitle{Bilinear mixed-effects models for dyadic data}.
\bjournal{Journal of the American Statistical Association}
\bvolume{100}
\bpages{286--295}.
\end{barticle}
\endbibitem

\bibitem[\protect\citeauthoryear{Hoff}{2007}]{hoff2007modeling}
\begin{barticle}[author]
\bauthor{\bsnm{Hoff},~\bfnm{Peter}\binits{P.}}
(\byear{2007}).
\btitle{Modeling homophily and stochastic equivalence in symmetric relational data}.
\bjournal{Advances in Neural Information Processing Systems}
\bvolume{20}.
\end{barticle}
\endbibitem

\bibitem[\protect\citeauthoryear{Hoff}{2011}]{hoff2011separable}
\begin{barticle}[author]
\bauthor{\bsnm{Hoff},~\bfnm{Peter~D}\binits{P.~D.}}
(\byear{2011}).
\btitle{Separable covariance arrays via the Tucker product, with applications to multivariate relational data}.
\bjournal{Bayesian Analysis}
\bvolume{6}
\bpages{179--196}.
\end{barticle}
\endbibitem

\bibitem[\protect\citeauthoryear{Hoff}{2021}]{Hoff2021Additive}
\begin{barticle}[author]
\bauthor{\bsnm{Hoff},~\bfnm{Peter}\binits{P.}}
(\byear{2021}).
\btitle{Additive and multiplicative effects network models}.
\bjournal{Statistical Science}
\bvolume{36}
\bpages{34--50}.
\end{barticle}
\endbibitem

\bibitem[\protect\citeauthoryear{Hoover}{1979}]{hoover1979relations}
\begin{barticle}[author]
\bauthor{\bsnm{Hoover},~\bfnm{Douglas~N}\binits{D.~N.}}
(\byear{1979}).
\btitle{Relations on probability spaces and arrays of random variables}.
\bjournal{Preprint, Institute for Advanced Study}.
\end{barticle}
\endbibitem

\bibitem[\protect\citeauthoryear{Hunter, Krivitsky and Schweinberger}{2012}]{hunter2012computational}
\begin{barticle}[author]
\bauthor{\bsnm{Hunter},~\bfnm{David~R}\binits{D.~R.}}, \bauthor{\bsnm{Krivitsky},~\bfnm{Pavel~N}\binits{P.~N.}} \AND \bauthor{\bsnm{Schweinberger},~\bfnm{Michael}\binits{M.}}
(\byear{2012}).
\btitle{Computational statistical methods for social network models}.
\bjournal{Journal of Computational and Graphical Statistics}
\bvolume{21}
\bpages{856--882}.
\end{barticle}
\endbibitem

\bibitem[\protect\citeauthoryear{Johnsson and Moon}{2021}]{johnsson2021estimation}
\begin{barticle}[author]
\bauthor{\bsnm{Johnsson},~\bfnm{Ida}\binits{I.}} \AND \bauthor{\bsnm{Moon},~\bfnm{Hyungsik~Roger}\binits{H.~R.}}
(\byear{2021}).
\btitle{Estimation of peer effects in endogenous social networks: Control function approach}.
\bjournal{The Review of Economics and Statistics}
\bvolume{103}
\bpages{328--345}.
\end{barticle}
\endbibitem

\bibitem[\protect\citeauthoryear{Kenny}{1988}]{kenny1988interpersonal}
\begin{barticle}[author]
\bauthor{\bsnm{Kenny},~\bfnm{David~A}\binits{D.~A.}}
(\byear{1988}).
\btitle{Interpersonal perception: A social relations analysis}.
\bjournal{Journal of Social and Personal Relationships}
\bvolume{5}
\bpages{247--261}.
\end{barticle}
\endbibitem

\bibitem[\protect\citeauthoryear{Kenny, Kashy and Cook}{2006}]{kenny2006analysis}
\begin{bbook}[author]
\bauthor{\bsnm{Kenny},~\bfnm{David~A}\binits{D.~A.}}, \bauthor{\bsnm{Kashy},~\bfnm{Deborah~A}\binits{D.~A.}} \AND \bauthor{\bsnm{Cook},~\bfnm{William~L}\binits{W.~L.}}
(\byear{2006}).
\btitle{Dyadic Data Analysis}.
\bpublisher{Guilford Press}.
\end{bbook}
\endbibitem

\bibitem[\protect\citeauthoryear{Kenny and La~Voie}{1984}]{kenny1984srm}
\begin{barticle}[author]
\bauthor{\bsnm{Kenny},~\bfnm{David~A}\binits{D.~A.}} \AND \bauthor{\bsnm{La~Voie},~\bfnm{Lawrence}\binits{L.}}
(\byear{1984}).
\btitle{The social relations model}.
\bjournal{Advances in Experimental Social Psychology}
\bvolume{18}
\bpages{141--182}.
\end{barticle}
\endbibitem

\bibitem[\protect\citeauthoryear{Kluger et~al.}{2021}]{kluger2021dyadic}
\begin{barticle}[author]
\bauthor{\bsnm{Kluger},~\bfnm{Avraham~N}\binits{A.~N.}}, \bauthor{\bsnm{Malloy},~\bfnm{Thomas~E}\binits{T.~E.}}, \bauthor{\bsnm{Pery},~\bfnm{Sarit}\binits{S.}}, \bauthor{\bsnm{Itzchakov},~\bfnm{Guy}\binits{G.}}, \bauthor{\bsnm{Castro},~\bfnm{Dotan~R}\binits{D.~R.}}, \bauthor{\bsnm{Lipetz},~\bfnm{Liora}\binits{L.}}, \bauthor{\bsnm{Sela},~\bfnm{Yaron}\binits{Y.}}, \bauthor{\bsnm{Turjeman-Levi},~\bfnm{Yaara}\binits{Y.}}, \bauthor{\bsnm{Lehmann},~\bfnm{Michal}\binits{M.}}, \bauthor{\bsnm{New},~\bfnm{Malki}\binits{M.}} \betal{et~al.}
(\byear{2021}).
\btitle{Dyadic listening in teams: Social relations model}.
\bjournal{Applied Psychology}
\bvolume{70}
\bpages{1045--1099}.
\end{barticle}
\endbibitem

\bibitem[\protect\citeauthoryear{Korolyuk and Borovskich}{2013}]{korolyuk2013theory}
\begin{bbook}[author]
\bauthor{\bsnm{Korolyuk},~\bfnm{Vladimir~S}\binits{V.~S.}} \AND \bauthor{\bsnm{Borovskich},~\bfnm{Yu~V}\binits{Y.~V.}}
(\byear{2013}).
\btitle{Theory of U-statistics}
\bvolume{273}.
\bpublisher{Springer Science \& Business Media}.
\end{bbook}
\endbibitem

\bibitem[\protect\citeauthoryear{Koster and Leckie}{2014}]{koster2014food}
\begin{barticle}[author]
\bauthor{\bsnm{Koster},~\bfnm{Jeremy~M}\binits{J.~M.}} \AND \bauthor{\bsnm{Leckie},~\bfnm{George}\binits{G.}}
(\byear{2014}).
\btitle{Food sharing networks in lowland Nicaragua: an application of the social relations model to count data}.
\bjournal{Social Networks}
\bvolume{38}
\bpages{100--110}.
\end{barticle}
\endbibitem

\bibitem[\protect\citeauthoryear{Lashley and Bond~Jr}{1997}]{lashley1997significance}
\begin{barticle}[author]
\bauthor{\bsnm{Lashley},~\bfnm{Brian~R}\binits{B.~R.}} \AND \bauthor{\bsnm{Bond~Jr},~\bfnm{Charles~F}\binits{C.~F.}}
(\byear{1997}).
\btitle{Significance testing for round robin data.}
\bjournal{Psychological Methods}
\bvolume{2}
\bpages{278--291}.
\end{barticle}
\endbibitem

\bibitem[\protect\citeauthoryear{Li and Loken}{2002}]{li2002unified}
\begin{barticle}[author]
\bauthor{\bsnm{Li},~\bfnm{Heng}\binits{H.}} \AND \bauthor{\bsnm{Loken},~\bfnm{Eric}\binits{E.}}
(\byear{2002}).
\btitle{A unified theory of statistical analysis and inference for variance component models for dyadic data}.
\bjournal{Statistica Sinica}
\bpages{519--535}.
\end{barticle}
\endbibitem

\bibitem[\protect\citeauthoryear{Liu and Xie}{2020}]{liu2020cauchy}
\begin{barticle}[author]
\bauthor{\bsnm{Liu},~\bfnm{Yaowu}\binits{Y.}} \AND \bauthor{\bsnm{Xie},~\bfnm{Jun}\binits{J.}}
(\byear{2020}).
\btitle{Cauchy combination test: a powerful test with analytic p-value calculation under arbitrary dependency structures}.
\bjournal{Journal of the American Statistical Association}
\bvolume{115}
\bpages{393--402}.
\end{barticle}
\endbibitem

\bibitem[\protect\citeauthoryear{Liu, Yu and Li}{2022}]{liu2022multiple}
\begin{barticle}[author]
\bauthor{\bsnm{Liu},~\bfnm{Wanjun}\binits{W.}}, \bauthor{\bsnm{Yu},~\bfnm{Xiufan}\binits{X.}} \AND \bauthor{\bsnm{Li},~\bfnm{Runze}\binits{R.}}
(\byear{2022}).
\btitle{Multiple-splitting projection test for high-dimensional mean vectors}.
\bjournal{The Journal of Machine Learning Research}
\bvolume{23}
\bpages{3091--3117}.
\end{barticle}
\endbibitem

\bibitem[\protect\citeauthoryear{L{\"u}dtke et~al.}{2013}]{ludtke2013general}
\begin{barticle}[author]
\bauthor{\bsnm{L{\"u}dtke},~\bfnm{Oliver}\binits{O.}}, \bauthor{\bsnm{Robitzsch},~\bfnm{Alexander}\binits{A.}}, \bauthor{\bsnm{Kenny},~\bfnm{David~A}\binits{D.~A.}} \AND \bauthor{\bsnm{Trautwein},~\bfnm{Ulrich}\binits{U.}}
(\byear{2013}).
\btitle{A general and flexible approach to estimating the social relations model using Bayesian methods.}
\bjournal{Psychological Methods}
\bvolume{18}
\bpages{101--119}.
\end{barticle}
\endbibitem

\bibitem[\protect\citeauthoryear{Meagher et~al.}{2021}]{meagher2021intellectual}
\begin{barticle}[author]
\bauthor{\bsnm{Meagher},~\bfnm{Benjamin~R}\binits{B.~R.}}, \bauthor{\bsnm{Leman},~\bfnm{Joseph~C}\binits{J.~C.}}, \bauthor{\bsnm{Heidenga},~\bfnm{Caitlyn~A}\binits{C.~A.}}, \bauthor{\bsnm{Ringquist},~\bfnm{Michala~R}\binits{M.~R.}} \AND \bauthor{\bsnm{Rowatt},~\bfnm{Wade~C}\binits{W.~C.}}
(\byear{2021}).
\btitle{Intellectual humility in conversation: Distinct behavioral indicators of self and peer ratings}.
\bjournal{The Journal of Positive Psychology}
\bvolume{16}
\bpages{417--429}.
\end{barticle}
\endbibitem

\bibitem[\protect\citeauthoryear{MoGinley and Sibson}{1975}]{moginley1975dissociated}
\begin{binproceedings}[author]
\bauthor{\bsnm{MoGinley},~\bfnm{WG}\binits{W.}} \AND \bauthor{\bsnm{Sibson},~\bfnm{Robin}\binits{R.}}
(\byear{1975}).
\btitle{Dissociated random variables}.
In \bbooktitle{Mathematical Proceedings of the Cambridge Philosophical Society}
\bvolume{77}
\bpages{185--188}.
\bpublisher{Cambridge University Press}.
\end{binproceedings}
\endbibitem

\bibitem[\protect\citeauthoryear{Motten and Stone}{2000}]{motten2000heritability}
\begin{barticle}[author]
\bauthor{\bsnm{Motten},~\bfnm{Alexander~F}\binits{A.~F.}} \AND \bauthor{\bsnm{Stone},~\bfnm{Judy~L}\binits{J.~L.}}
(\byear{2000}).
\btitle{Heritability of stigma position and the effect of stigma-anther separation on outcrossing in a predominantly self-fertilizing weed, \emph{Datura} \emph{stramonium} (Solanaceae)}.
\bjournal{American Journal of Botany}
\bvolume{87}
\bpages{339--347}.
\end{barticle}
\endbibitem

\bibitem[\protect\citeauthoryear{Nestler}{2016}]{nestler2016restricted}
\begin{barticle}[author]
\bauthor{\bsnm{Nestler},~\bfnm{Steffen}\binits{S.}}
(\byear{2016}).
\btitle{Restricted maximum likelihood estimation for parameters of the social relations model}.
\bjournal{Psychometrika}
\bvolume{81}
\bpages{1098--1117}.
\end{barticle}
\endbibitem

\bibitem[\protect\citeauthoryear{Nestler, L{\"u}dtke and Robitzsch}{2020}]{nestler2020maximum}
\begin{barticle}[author]
\bauthor{\bsnm{Nestler},~\bfnm{Steffen}\binits{S.}}, \bauthor{\bsnm{L{\"u}dtke},~\bfnm{Oliver}\binits{O.}} \AND \bauthor{\bsnm{Robitzsch},~\bfnm{Alexander}\binits{A.}}
(\byear{2020}).
\btitle{Maximum likelihood estimation of a social relations structural equation model}.
\bjournal{Psychometrika}
\bvolume{85}
\bpages{870--889}.
\end{barticle}
\endbibitem

\bibitem[\protect\citeauthoryear{Onnela et~al.}{2007}]{onnela2007structure}
\begin{barticle}[author]
\bauthor{\bsnm{Onnela},~\bfnm{J-P}\binits{J.-P.}}, \bauthor{\bsnm{Saram{\"a}ki},~\bfnm{Jari}\binits{J.}}, \bauthor{\bsnm{Hyv{\"o}nen},~\bfnm{Jorkki}\binits{J.}}, \bauthor{\bsnm{Szab{\'o}},~\bfnm{Gy{\"o}rgy}\binits{G.}}, \bauthor{\bsnm{Lazer},~\bfnm{David}\binits{D.}}, \bauthor{\bsnm{Kaski},~\bfnm{Kimmo}\binits{K.}}, \bauthor{\bsnm{Kert{\'e}sz},~\bfnm{J{\'a}nos}\binits{J.}} \AND \bauthor{\bsnm{Barab{\'a}si},~\bfnm{A-L}\binits{A.-L.}}
(\byear{2007}).
\btitle{Structure and tie strengths in mobile communication networks}.
\bjournal{Proceedings of the National Academy of Sciences}
\bvolume{104}
\bpages{7332--7336}.
\end{barticle}
\endbibitem

\bibitem[\protect\citeauthoryear{Opsahl and Panzarasa}{2009}]{opsahl2009clustering}
\begin{barticle}[author]
\bauthor{\bsnm{Opsahl},~\bfnm{Tore}\binits{T.}} \AND \bauthor{\bsnm{Panzarasa},~\bfnm{Pietro}\binits{P.}}
(\byear{2009}).
\btitle{Clustering in weighted networks}.
\bjournal{Social Networks}
\bvolume{31}
\bpages{155--163}.
\end{barticle}
\endbibitem

\bibitem[\protect\citeauthoryear{Owen and Eckles}{2012}]{Art2012a}
\begin{barticle}[author]
\bauthor{\bsnm{Owen},~\bfnm{Art~B.}\binits{A.~B.}} \AND \bauthor{\bsnm{Eckles},~\bfnm{Dean}\binits{D.}}
(\byear{2012}).
\btitle{{Bootstrapping data arrays of arbitrary order}}.
\bjournal{The Annals of Applied Statistics}
\bvolume{6}
\bpages{895 -- 927}.
\bdoi{10.1214/12-AOAS547}
\end{barticle}
\endbibitem

\bibitem[\protect\citeauthoryear{Schweinberger and Stewart}{2020}]{Schweinberger2020Concentration}
\begin{barticle}[author]
\bauthor{\bsnm{Schweinberger},~\bfnm{Michael}\binits{M.}} \AND \bauthor{\bsnm{Stewart},~\bfnm{Jonathan}\binits{J.}}
(\byear{2020}).
\btitle{{Concentration and consistency results for canonical and curved exponential-family models of random graphs}}.
\bjournal{The Annals of Statistics}
\bvolume{48}
\bpages{374--396}.
\bdoi{10.1214/19-AOS1810}
\end{barticle}
\endbibitem

\bibitem[\protect\citeauthoryear{Serfling}{1980}]{serfling1980approximation}
\begin{barticle}[author]
\bauthor{\bsnm{Serfling},~\bfnm{R}\binits{R.}}
(\byear{1980}).
\btitle{Approximation Theorems of Mathematical Statistics}.
\bjournal{Wiley Series in Probability and Statistics}.
\end{barticle}
\endbibitem

\bibitem[\protect\citeauthoryear{Shao, Xia and Zhang}{2023}]{shao2023U-statistics}
\begin{barticle}[author]
\bauthor{\bsnm{Shao},~\bfnm{Meijia}\binits{M.}}, \bauthor{\bsnm{Xia},~\bfnm{Dong}\binits{D.}} \AND \bauthor{\bsnm{Zhang},~\bfnm{Yuan}\binits{Y.}}
(\byear{2023}).
\btitle{U-Statistic reduction: higher-order accurate risk control and statistical-computational trade-off, with application to network method-of-moments}.
\bjournal{arXiv preprint arXiv:2306.03793}.
\end{barticle}
\endbibitem

\bibitem[\protect\citeauthoryear{Shao et~al.}{2022}]{shao2022higher}
\begin{barticle}[author]
\bauthor{\bsnm{Shao},~\bfnm{Meijia}\binits{M.}}, \bauthor{\bsnm{Xia},~\bfnm{Dong}\binits{D.}}, \bauthor{\bsnm{Zhang},~\bfnm{Yuan}\binits{Y.}}, \bauthor{\bsnm{Wu},~\bfnm{Qiong}\binits{Q.}} \AND \bauthor{\bsnm{Chen},~\bfnm{Shuo}\binits{S.}}
(\byear{2022}).
\btitle{Higher-order accurate two-sample network inference and network hashing}.
\bjournal{arXiv preprint arXiv:2208.07573}.
\end{barticle}
\endbibitem

\bibitem[\protect\citeauthoryear{Silva and Tenreyro}{2006}]{silva2006log}
\begin{barticle}[author]
\bauthor{\bsnm{Silva},~\bfnm{JMC~Santos}\binits{J.~S.}} \AND \bauthor{\bsnm{Tenreyro},~\bfnm{Silvana}\binits{S.}}
(\byear{2006}).
\btitle{The log of gravity}.
\bjournal{The Review of Economics and Statistics}
\bvolume{88}
\bpages{641--658}.
\end{barticle}
\endbibitem

\bibitem[\protect\citeauthoryear{Silva and Tenreyro}{2010}]{silva2010currency}
\begin{barticle}[author]
\bauthor{\bsnm{Silva},~\bfnm{JMC~Santos}\binits{J.~S.}} \AND \bauthor{\bsnm{Tenreyro},~\bfnm{Silvana}\binits{S.}}
(\byear{2010}).
\btitle{Currency unions in prospect and retrospect}.
\bjournal{Annual Review of Economics}
\bvolume{2}
\bpages{51--74}.
\end{barticle}
\endbibitem

\bibitem[\protect\citeauthoryear{Silverman}{1976}]{Silverman1976weaklyExch}
\begin{barticle}[author]
\bauthor{\bsnm{Silverman},~\bfnm{Bernard~W}\binits{B.~W.}}
(\byear{1976}).
\btitle{Limit theorems for dissociated random variables}.
\bjournal{Advances in Applied Probability}
\bvolume{8}
\bpages{806--819}.
\end{barticle}
\endbibitem

\bibitem[\protect\citeauthoryear{Simpson et~al.}{2013}]{simpson2013permutation}
\begin{barticle}[author]
\bauthor{\bsnm{Simpson},~\bfnm{Sean~L}\binits{S.~L.}}, \bauthor{\bsnm{Lyday},~\bfnm{Robert~G}\binits{R.~G.}}, \bauthor{\bsnm{Hayasaka},~\bfnm{Satoru}\binits{S.}}, \bauthor{\bsnm{Marsh},~\bfnm{Anthony~P}\binits{A.~P.}} \AND \bauthor{\bsnm{Laurienti},~\bfnm{Paul~J}\binits{P.~J.}}
(\byear{2013}).
\btitle{A permutation testing framework to compare groups of brain networks}.
\bjournal{Frontiers in Computational Neuroscience}
\bvolume{7}
\bpages{171}.
\end{barticle}
\endbibitem

\bibitem[\protect\citeauthoryear{Snijders}{2002}]{Snijders2002MCMCeponential}
\begin{barticle}[author]
\bauthor{\bsnm{Snijders},~\bfnm{Tom~AB}\binits{T.~A.}}
(\byear{2002}).
\btitle{Markov chain Monte Carlo estimation of exponential random graph models}.
\bjournal{Journal of Social Structure}
\bvolume{3}
\bpages{1--40}.
\end{barticle}
\endbibitem

\bibitem[\protect\citeauthoryear{Snijders}{2011}]{snijders2011statistical}
\begin{barticle}[author]
\bauthor{\bsnm{Snijders},~\bfnm{Tom~AB}\binits{T.~A.}}
(\byear{2011}).
\btitle{Statistical models for social networks}.
\bjournal{Annual Review of Sociology}
\bvolume{37}
\bpages{131--153}.
\end{barticle}
\endbibitem

\bibitem[\protect\citeauthoryear{Snijders and Kenny}{1999}]{snijders1999srm}
\begin{barticle}[author]
\bauthor{\bsnm{Snijders},~\bfnm{Tom~AB}\binits{T.~A.}} \AND \bauthor{\bsnm{Kenny},~\bfnm{David~A}\binits{D.~A.}}
(\byear{1999}).
\btitle{The social relations model for family data: A multilevel approach}.
\bjournal{Personal Relationships}
\bvolume{6}
\bpages{471--486}.
\end{barticle}
\endbibitem

\bibitem[\protect\citeauthoryear{Snijders et~al.}{2006}]{Snijders2006eponentialGraphs}
\begin{barticle}[author]
\bauthor{\bsnm{Snijders},~\bfnm{Tom~AB}\binits{T.~A.}}, \bauthor{\bsnm{Pattison},~\bfnm{Philippa~E}\binits{P.~E.}}, \bauthor{\bsnm{Robins},~\bfnm{Garry~L}\binits{G.~L.}} \AND \bauthor{\bsnm{Handcock},~\bfnm{Mark~S}\binits{M.~S.}}
(\byear{2006}).
\btitle{New specifications for exponential random graph models}.
\bjournal{Sociological Methodology}
\bvolume{36}
\bpages{99--153}.
\end{barticle}
\endbibitem

\bibitem[\protect\citeauthoryear{Squartini et~al.}{2013}]{Squartini2013reciprocityNetwork}
\begin{barticle}[author]
\bauthor{\bsnm{Squartini},~\bfnm{Tiziano}\binits{T.}}, \bauthor{\bsnm{Picciolo},~\bfnm{Francesco}\binits{F.}}, \bauthor{\bsnm{Ruzzenenti},~\bfnm{Franco}\binits{F.}} \AND \bauthor{\bsnm{Garlaschelli},~\bfnm{Diego}\binits{D.}}
(\byear{2013}).
\btitle{Reciprocity of weighted networks}.
\bjournal{Scientific Reports}
\bvolume{3}
\bpages{1--9}.
\end{barticle}
\endbibitem

\bibitem[\protect\citeauthoryear{Tinbergen}{1962}]{tinbergen1962shaping}
\begin{bbook}[author]
\bauthor{\bsnm{Tinbergen},~\bfnm{Jan}\binits{J.}}
(\byear{1962}).
\btitle{Shaping the World Economy, Suggestions for an International Economic Policy}.
\bpublisher{Twentieth Century Fund}.
\end{bbook}
\endbibitem

\bibitem[\protect\citeauthoryear{Underwood, Elliott and Cucuringu}{2020}]{underwood2020motif}
\begin{barticle}[author]
\bauthor{\bsnm{Underwood},~\bfnm{William~G}\binits{W.~G.}}, \bauthor{\bsnm{Elliott},~\bfnm{Andrew}\binits{A.}} \AND \bauthor{\bsnm{Cucuringu},~\bfnm{Mihai}\binits{M.}}
(\byear{2020}).
\btitle{Motif-based spectral clustering of weighted directed networks}.
\bjournal{Applied Network Science}
\bvolume{5}
\bpages{1--41}.
\end{barticle}
\endbibitem

\bibitem[\protect\citeauthoryear{Van~Borkulo et~al.}{2022}]{van2022comparing}
\begin{barticle}[author]
\bauthor{\bsnm{Van~Borkulo},~\bfnm{Claudia~D}\binits{C.~D.}}, \bauthor{\bparticle{van} \bsnm{Bork},~\bfnm{Riet}\binits{R.}}, \bauthor{\bsnm{Boschloo},~\bfnm{Lynn}\binits{L.}}, \bauthor{\bsnm{Kossakowski},~\bfnm{Jolanda~J}\binits{J.~J.}}, \bauthor{\bsnm{Tio},~\bfnm{Pia}\binits{P.}}, \bauthor{\bsnm{Schoevers},~\bfnm{Robert~A}\binits{R.~A.}}, \bauthor{\bsnm{Borsboom},~\bfnm{Denny}\binits{D.}} \AND \bauthor{\bsnm{Waldorp},~\bfnm{Lourens~J}\binits{L.~J.}}
(\byear{2022}).
\btitle{Comparing network structures on three aspects: A permutation test}.
\bjournal{Psychological Methods}.
\end{barticle}
\endbibitem

\bibitem[\protect\citeauthoryear{Van~der Vaart}{2000}]{van2000asymptotic}
\begin{bbook}[author]
\bauthor{\bparticle{Van~der} \bsnm{Vaart},~\bfnm{Aad~W}\binits{A.~W.}}
(\byear{2000}).
\btitle{Asymptotic Statistics}.
\bpublisher{Cambridge University Press}.
\end{bbook}
\endbibitem

\bibitem[\protect\citeauthoryear{Vershynin}{2018}]{vershynin2018high}
\begin{bbook}[author]
\bauthor{\bsnm{Vershynin},~\bfnm{Roman}\binits{R.}}
(\byear{2018}).
\btitle{High-Dimensional Probability: An Introduction with Applications in Data Science}
\bvolume{47}.
\bpublisher{Cambridge University Press}.
\end{bbook}
\endbibitem

\bibitem[\protect\citeauthoryear{Ward and Hoff}{2007}]{ward2007persistent}
\begin{barticle}[author]
\bauthor{\bsnm{Ward},~\bfnm{Michael~D}\binits{M.~D.}} \AND \bauthor{\bsnm{Hoff},~\bfnm{Peter~D}\binits{P.~D.}}
(\byear{2007}).
\btitle{Persistent patterns of international commerce}.
\bjournal{Journal of Peace Research}
\bvolume{44}
\bpages{157--175}.
\end{barticle}
\endbibitem

\bibitem[\protect\citeauthoryear{Warner, Kenny and Stoto}{1979}]{Warner1979roundRobin}
\begin{barticle}[author]
\bauthor{\bsnm{Warner},~\bfnm{Rebecca~M}\binits{R.~M.}}, \bauthor{\bsnm{Kenny},~\bfnm{David~A}\binits{D.~A.}} \AND \bauthor{\bsnm{Stoto},~\bfnm{Michael}\binits{M.}}
(\byear{1979}).
\btitle{A new round robin analysis of variance for social interaction data}.
\bjournal{Journal of Personality and Social Psychology}
\bvolume{37}
\bpages{1742--1757}.
\end{barticle}
\endbibitem

\bibitem[\protect\citeauthoryear{Weber}{1981}]{weber1981incomplete}
\begin{barticle}[author]
\bauthor{\bsnm{Weber},~\bfnm{NC}\binits{N.}}
(\byear{1981}).
\btitle{Incomplete degenerate U-statistics}.
\bjournal{Scandinavian Journal of Statistics}
\bpages{120--123}.
\end{barticle}
\endbibitem

\bibitem[\protect\citeauthoryear{Westveld and Hoff}{2011}]{westveld2011mem}
\begin{barticle}[author]
\bauthor{\bsnm{Westveld},~\bfnm{Anton~H}\binits{A.~H.}} \AND \bauthor{\bsnm{Hoff},~\bfnm{Peter~D}\binits{P.~D.}}
(\byear{2011}).
\btitle{A mixed effects model for longitudinal relational and network data, with applications to international trade and conflict}.
\bjournal{The Annals of Applied Statistics}
\bpages{843--872}.
\end{barticle}
\endbibitem

\bibitem[\protect\citeauthoryear{Wong}{1982}]{Wong1982roundRobin}
\begin{barticle}[author]
\bauthor{\bsnm{Wong},~\bfnm{George~Y}\binits{G.~Y.}}
(\byear{1982}).
\btitle{Round robin analysis of variance via maximum likelihood}.
\bjournal{Journal of the American Statistical Association}
\bvolume{77}
\bpages{714--724}.
\end{barticle}
\endbibitem

\bibitem[\protect\citeauthoryear{Yao, Zhang and Shao}{2018}]{yao2018testing}
\begin{barticle}[author]
\bauthor{\bsnm{Yao},~\bfnm{Shun}\binits{S.}}, \bauthor{\bsnm{Zhang},~\bfnm{Xianyang}\binits{X.}} \AND \bauthor{\bsnm{Shao},~\bfnm{Xiaofeng}\binits{X.}}
(\byear{2018}).
\btitle{Testing mutual independence in high dimension via distance covariance}.
\bjournal{Journal of the Royal Statistical Society Series B: Statistical Methodology}
\bvolume{80}
\bpages{455--480}.
\end{barticle}
\endbibitem

\bibitem[\protect\citeauthoryear{Yuan and Qu}{2021}]{yuan2021community}
\begin{barticle}[author]
\bauthor{\bsnm{Yuan},~\bfnm{Yubai}\binits{Y.}} \AND \bauthor{\bsnm{Qu},~\bfnm{Annie}\binits{A.}}
(\byear{2021}).
\btitle{Community detection with dependent connectivity}.
\bjournal{The Annals of Statistics}
\bvolume{49}
\bpages{2378--2428}.
\end{barticle}
\endbibitem

\bibitem[\protect\citeauthoryear{Zeleneev}{2020}]{zeleneev2020identification}
\begin{barticle}[author]
\bauthor{\bsnm{Zeleneev},~\bfnm{Andrei}\binits{A.}}
(\byear{2020}).
\btitle{Identification and estimation of network models with nonparametric unobserved heterogeneity}.
\bjournal{Department of Economics, Princeton University}.
\end{barticle}
\endbibitem

\bibitem[\protect\citeauthoryear{Zhang, Levina and Zhu}{2017}]{zhang2017estimating}
\begin{barticle}[author]
\bauthor{\bsnm{Zhang},~\bfnm{Yuan}\binits{Y.}}, \bauthor{\bsnm{Levina},~\bfnm{Elizaveta}\binits{E.}} \AND \bauthor{\bsnm{Zhu},~\bfnm{Ji}\binits{J.}}
(\byear{2017}).
\btitle{Estimating network edge probabilities by neighbourhood smoothing}.
\bjournal{Biometrika}
\bvolume{104}
\bpages{771--783}.
\end{barticle}
\endbibitem

\bibitem[\protect\citeauthoryear{Zhang and Xia}{2022}]{Zhang2022edgeworth}
\begin{barticle}[author]
\bauthor{\bsnm{Zhang},~\bfnm{Yuan}\binits{Y.}} \AND \bauthor{\bsnm{Xia},~\bfnm{Dong}\binits{D.}}
(\byear{2022}).
\btitle{Edgeworth expansions for network moments}.
\bjournal{The Annals of Statistics}
\bvolume{50}
\bpages{726--753}.
\end{barticle}
\endbibitem

\bibitem[\protect\citeauthoryear{Zhang, Zhao and Zhou}{2021}]{zhang2021beauty}
\begin{barticle}[author]
\bauthor{\bsnm{Zhang},~\bfnm{Kai}\binits{K.}}, \bauthor{\bsnm{Zhao},~\bfnm{Zhigen}\binits{Z.}} \AND \bauthor{\bsnm{Zhou},~\bfnm{Wen}\binits{W.}}
(\byear{2021}).
\btitle{Beauty powered beast}.
\bjournal{arXiv preprint arXiv:2103.00674}.
\end{barticle}
\endbibitem

\bibitem[\protect\citeauthoryear{Zhang and Zhou}{2016}]{zhang2016minimax}
\begin{barticle}[author]
\bauthor{\bsnm{Zhang},~\bfnm{Anderson~Y}\binits{A.~Y.}} \AND \bauthor{\bsnm{Zhou},~\bfnm{Harrison~H}\binits{H.~H.}}
(\byear{2016}).
\btitle{Minimax Rates of Community Detection in Stochastic Block Models}.
\bjournal{The Annals of Statistics}
\bvolume{44}
\bpages{2252--2280}.
\end{barticle}
\endbibitem

\end{thebibliography}


\end{document}